\documentclass{aa} 
\usepackage{graphicx}
\usepackage{natbib}
\usepackage{multirow}
\usepackage{longtable}
\usepackage{lscape}
\usepackage{rotating}

\bibpunct{(}{)}{;}{a}{,}{,}

\usepackage{txfonts}
\usepackage{color}

\begin{document}

   \title{Gas phase Elemental abundances in Molecular cloudS (GEMS)}
   \subtitle{ IV. Observational results and statistical trends}

   \author{
    M. Rodr{\'{\i}}guez-Baras\inst{1}
    \and
     A.~Fuente\inst{1}
     \and
     P.~Rivi\'ere-Marichalar\inst{1}
    %
%     \and
   D. Navarro-Almaida\inst{1}
     \and
     P.~Caselli\inst{2}
     \and
     M.~Gerin\inst{3}     
     \and
     C.~Kramer\inst{4}          
     \and
    E.~Roueff\inst{5}       
    \and
   V.~Wakelam\inst{6}
    \and
   G. Esplugues\inst{1}
    \and
   S.~Garc\'{\i}a-Burillo\inst{1}
  \and
  R. Le Gal\inst{7}
   \and
   S. Spezzano\inst{2}
   \and
   T.~ Alonso-Albi \inst{1}
    \and         
    R.~ Bachiller \inst{1}
   \and      
   S.~Cazaux \inst{8}
   \and      
   B.~Commercon\inst{9}
   \and   
   J.~ R.~Goicoechea \inst{10}
   \and
    J.~C.~Loison\inst{11}
   \and
    S.~P. Trevi\~no-Morales\inst{12}
      \and
     O.~Roncero\inst{10}
    \and
    I.~Jim\'enez-Serra\inst{13}
 \and
 J. Laas\inst{2}
  \and
    A.~Hacar \inst{14}
     \and
    J.~Kirk\inst{15}
    \and
    V.~Lattanzi\inst{2}
     \and
     R.~ Mart\'{\i}n-Dom\'enech\inst{7}
% 
 %    \and
     G.~Mu\~noz-Caro\inst{13}
     \and
    J.~E.~Pineda\inst{2}
    \and
    B.~Tercero\inst{1,16}
    \and
    D.~Ward-Thompson\inst{15}             
     \and
    M.~Tafalla\inst{1}
     \and
     N.~Marcelino\inst{10}
     \and
    J.~Malinen\inst{17,18}
    \and
   R.~Friesen\inst{19}
   \and
   B.~M.~Giuliano\inst{2}
    }
             
   \institute{Observatorio Astron\'omico Nacional (OAN), Alfonso XII, 3,  28014, Madrid, Spain
    \and
    Centre for Astrochemical Studies, Max-Planck-Institute for Extraterrestrial Physics, Giessenbachstrasse 1, 85748, Garching, Germany    
    \and
   Observatoire de Paris, PSL Research University, CNRS, \'Ecole Normale Sup\'erieure, Sorbonne Universit\'es, UPMC Univ. Paris 06, 75005, Paris, France
   \and
  Institut de Radioastronomie Millimétrique, 300 rue de la Piscine, Domaine Universitaire, 38406 Saint Martin d'Hères, France
   \and
  LERMA, Observatoire de PARIS, PSL Research University, CNRS, Sorbonne Universit\'e, 92190 Meudon, France
  \and
  Laboratoire d'astrophysique de Bordeaux, Univ. Bordeaux, CNRS, B18N, all\'ee Geoffroy Saint-Hilaire, 33615 Pessac, France 
  \and
  Harvard-Smithsonian Center for Astrophysics, 60 Garden St., Cambridge, MA 02138, USA
  \and
  Faculty of Aerospace Engineering, Delft University of Technology, Delft, The Netherlands ; University of Leiden, P.O. Box 9513, NL, 2300 RA, Leiden, The Netherlands 
  \and
  \'Ecole Normale Sup\'erieure de Lyon, CRAL, UMR CNRS 5574, Universit\'e Lyon I, 46 All\'ee d'Italie, 69364, Lyon Cedex 07, France 
  \and
  Instituto de F\'{\i}sica Fundamental (CSIC), Calle Serrano 123, 28006, Madrid, Spain
  \and
  Institut des Sciences Mol\'eculaires (ISM), CNRS, Univ. Bordeaux, 351 cours de la Lib\'eration, F-33400, Talence, France
  \and
  Chalmers University of Technology, Department of Space, Earth and Environment, SE-412 93 Gothenburg, Sweden
  \and
 Centro de Astrobiolog\'{\i}a (CSIC-INTA), Ctra. de Ajalvir, km 4, Torrej\'on de Ardoz, 28850, Madrid, Spain
 \and
 University of Vienna, Department of Astrophysics, Tuerkenschanzstrasse 17, 1180, Vienna
 \and
 Jeremiah Horrocks Institute, University of Central Lancashire, Preston PR1 2HE, UK
 \and
  Observatorio de Yebes (IGN). Cerro de la Palera s/n, 19141 Yebes, Spain
\and
 Department of Physics, University of Helsinki, PO Box 64, 00014 Helsinki, Finland
 \and
 Institute of Physics I, University of Cologne, Cologne, Germany
 \and
 National Radio Astronomy Observatory, 520 Edgemont Rd., Charlottesville VA 22901, USA 
 }
   
 \abstract  {
 Gas phase Elemental abundances in Molecular CloudS (GEMS) is an IRAM 30m Large Program designed to provide estimates of the S, C, N, and O depletions and gas ionization degree, X(e$^-$), in a selected set of star-forming filaments of Taurus, Perseus, and Orion. Our immediate goal is to build up a complete and large database of molecular abundances that can serve as an observational basis for estimating X(e$^-$) and the C, O, N, and S depletions through chemical modeling. 
We observed and derived the abundances of 14 species ($^{13}$CO, C$^{18}$O, HCO$^+$, H$^{13}$CO$^+$, HC$^{18}$O$^+$, HCN, H$^{13}$CN, HNC, HCS$^+$, CS, SO, $^{34}$SO, H$_2$S, and OCS) in 244 positions, covering the A$_V$$\sim$3 to $\sim$100~mag, $n(H_2)$$\sim$ a few 10$^3$ to 10$^6$~cm$^{-3}$, and T$_{k}\sim$10 to $\sim$30~K ranges in these clouds, and avoiding protostars, HII regions, and bipolar outflows. A statistical analysis is carried out in order to identify general trends between different species and with physical parameters. 
Relations between molecules reveal strong linear correlations which define three different families of species: (1) $^{13}$CO and C$^{18}$O isotopologs;  (2) H$^{13}$CO$^+$, HC$^{18}$O$^+$, H$^{13}$CN, and HNC; and (3) the S-bearing molecules. The abundances of the CO isotopologs increase with the gas kinetic temperature until T$_K \sim$~15~K. For higher temperatures, the abundance remains constant with a scatter of a factor of $\sim$3. The abundances of H$^{13}$CO$^+$, HC$^{18}$O$^+$, H$^{13}$CN, and  HNC are well correlated with each other, and all of them decrease with molecular hydrogen density, following the law $\propto$ $n(H_2)^{-0.8\pm0.2}$. The abundances of S-bearing species also decrease with molecular hydrogen density at a rate of (S-bearing/H)$_{gas}$ $\propto$ $n(H_2)^{-0.6\pm0.1}$. The abundances of molecules belonging to groups 2 and 3 do not present any clear trend with gas temperature. At scales of molecular clouds, the C$^{18}$O abundance is the quantity that better correlates with the cloud mass. We discuss the utility of the $^{13}$CO/C$^{18}$O, HCO$^+$/H$^{13}$CO$^+$, and H$^{13}$CO$^+$/H$^{13}$CN abundance ratios as chemical diagnostics of star formation in external galaxies. }

   \keywords{Astrochemistry -- ISM: abundances -- ISM: molecules -- ISM: clouds --
   stars: formation -- galaxies: ISM}
   
 \maketitle  

%%%%%%%%%%%%%%%%%%%

\section{Introduction}

Gas chemistry plays a key role in the star formation process by regulating fundamental parameters such as the gas cooling rate and ionization fraction. Molecular clouds can contract and fragment because molecules cool the gas, thus diminishing the thermal support relative to self-gravity. The ionization fraction controls the coupling of magnetic fields with the gas, driving the dissipation of turbulence and angular momentum transfer, and therefore plays a crucial role in cloud collapse (isolated vs. clustered star formation) and the dynamics of accretion disks (see \citealp{Zhao2016, Padovani2013}). In the absence of other ionization agents (X-rays, UV photons, J-type shocks), the steady state ionization fraction is proportional to $\sqrt{\zeta _{H_2}/n}$, where $n$ is the molecular hydrogen density and $\zeta _{H_2}$ is the cosmic-ray ionization rate for H$_2$ molecules, which becomes the key parameter in the molecular cloud evolution \citep{Oppenheimer1974, Kee1989, Caselli2002}. The gas ionization fraction, X(e$^-$)=n(e$^-$)/n$_{\rm H}$, and molecular abundances depend on the elemental depletion factors \citep{Caselli1998}. In particular, carbon (C) is the main donor of electrons in the cloud surface region (A$_V <$ 2 mag), and because of its lower ionization potential and as long as it is not heavily depleted, sulfur (S) is the main donor in the $\sim$3.7$-$7 magnitudes range that encompasses a large fraction of the molecular cloud mass \citep{Goicoechea2006}. As CO and [CII] are the main coolants, depletions of C and O determine the gas cooling rate in molecular clouds. 

Elemental depletions also constitute a valuable piece of information for our understanding of the grain composition and evolution. For any given element, the missing atoms in the gas phase are presumed to be locked up in solids, that is, dust grains and/or icy mantles. Knowledge of elemental depletions can therefore be valuable in studying the changes in the dust grain composition across the cloud. Surface chemistry and the interchange of molecules between the solid and gas phases have a leading role in the chemical evolution of  gas from the diffuse cloud to the prestellar core phase.

\setcitestyle{notesep={; },round,aysep={},yysep={;}}

\begin{figure}
%\special{psfile=14905f4.eps hoffset=-20 voffset=-230 hscale=40 vscale=35 angle=-0}
%\vspace{8.0cm}
\includegraphics[angle=0,scale=.25]{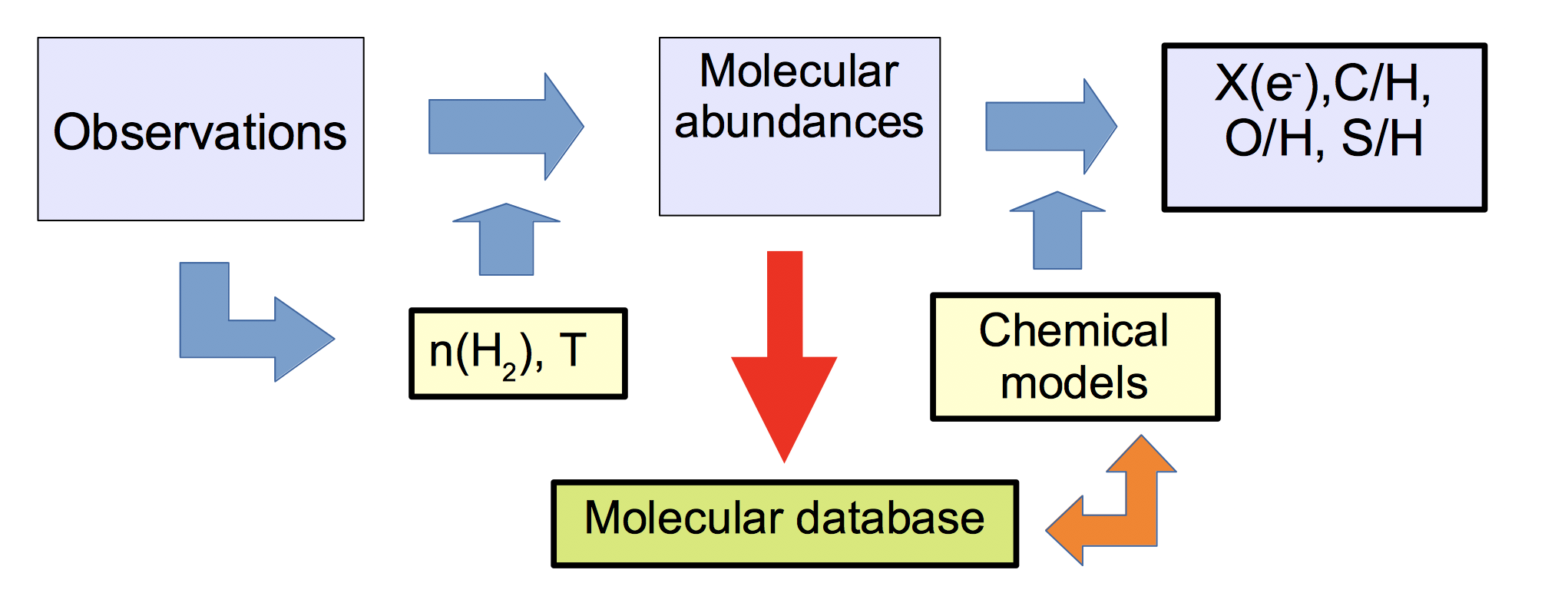}
\caption{ Block diagram of the GEMS project. }
\label{Fig: scheme}
%\vspace{-0.1cm}
\end{figure}

GEMS Gas phase Elemental abundances in Molecular CloudS (GEMS) is an IRAM 30m Large Program, the aim of which is to estimate the S, C, N, and O depletions and X(e$^-$) as a function of visual extinction in a selected set of prototypical star-forming filaments. Regions with different illumination are included in the sample in order to investigate the influence of UV radiation (photodissociation, ionization, photodesorption) and turbulence on these parameters, and eventually in the star formation history of the cloud. The depletion factor is defined as the ratio between the total (dust+gas) abundance of a given element and its abundance as observed in the gas phase. The determination of the depletion factor of a given element  can only been done directly through high-sensitivity observations of the main molecular reservoirs of each element and detailed chemical modeling of the secondary reservoirs \citep[][see a simple scheme in Fig. \ref{Fig: scheme}]{Fuente2019}. 

The first step to derive the elemental gas abundance of C, O, N, and S is to determine the abundances of the main reservoirs of the elements in the gas phase. Essentially, most of the carbon is locked in CO in dense cores and the C depletion is derived from the study of CO and its isotopologs. A significant fraction of C may be atomic (C, C$^+$) in the molecular gas surrounding the cores. The gas ionization fraction, X(e$^-$), can be derived from the [HCO$^+$]/[CO]  ratio, which can help to constrain the C depletion values through comparison with chemical models. The main reservoirs of nitrogen are supposed to be atomic nitrogen (N) and molecular nitrogen (N$_2$) which are not observable. The nitrogen abundance can be derived by applying a chemical model to fit the observed abundances of nitriles (HCN, HNC, CN)
\citep{Agundez2013, Legal2014}. The HCN abundance is also dependent on the amount of atomic C in gas phase, mainly on the C/O ratio \citep{Loison2014, Fuente2016}. The most abundant oxygenated molecules, O$_2$, H$_2$O, and OH, are difficult to observe in the millimeter domain and the oxygen depletion factor has to be derived indirectly from the C/O ratio. The CS/SO abundance ratio has been used as a proxy for the C/O ratio in different environments \citep{Fuente2016, Semenov2018}. The abundances of other species such as HCN and C$_2$H are also very sensitive to the C/O gas-phase ratio \citep{Loison2014, Miotello2019}.

Determining sulfur depletion is challenging. Sulfur is one of the most abundant elements in the Universe (S/H$\sim$1.5$\times$10$^{-5}$) \citep{Asplund2009} and plays a crucial role in biological systems on Earth, and so it is important to follow its chemical history in space (i.e., toward precursors of Solar System analogs). Surprisingly, S-bearing molecules are not as abundant as expected in the interstellar medium. Sulfur is thought to be depleted in molecular clouds by a factor up to 1000 compared to its estimated cosmic abundance \citep{Ruffle1999, Wakelam2004}. Because of the high hydrogen abundances and the mobility of hydrogen in the ice matrix, sulfur atoms in interstellar ice mantles are expected to preferentially form H$_2$S. There are only upper limits to the solid H$_2$S abundance (e.g., \citealp{JimenezEscobar2011}), and OCS is the only S-bearing molecule unambiguously detected in ice mantles because of its large band strength in the infrared \citep{Geballe1985,Palumbo1995} along with, tentatively, SO$_2$ \citep{Boogert1997}. One possibility is that most of the sulfur is locked in atomic sulfur in the gas phase. Also, sulfur could be in refractory allotropes, mainly S$_8$, as found theoretically by \citet{Shingledecker2020}, and previously observed in the laboratory by \citet{JimenezEscobar2012,JimenezEscobar2014}. The detection of sulfur allotropes in comets \citep{Calmonte2016} and of S$_2$H in gas phase \citep{Fuente2017} testify to the importance of these compounds.  Unfortunately, sulfur allotropes cannot be directly observed in the interstellar medium. Direct observation of the S atom is also difficult and, until now, S has only been detected in some bipolar outflows using the infrared space telescope Spitzer \citep{Anderson2013}. Our project includes a wide set of S-bearing species (CS, SO, H$_2$S, HCS$^+$, SO$_2$, OCS, and H$_2$CS) that are going to be used to constrain the sulfur depletion in our sample. To our knowledge, this will constitute the most complete database of S-bearing molecules in dark clouds so far. 

A detailed presentation of the project with a list of species and isotopologs observed was included in \citet{Fuente2019}. This paper presents the abundances of $^{13}$CO, C$^{18}$O, CS, C$^{34}$S, $^{13}$CS, H$_2$S, SO, $^{34}$SO, HCO$^+$, H$^{13}$CO$^+$, HC$^{18}$O$^+$, HCN, H$^{13}$CN, HNC, HCS$^+$, and OCS towards the whole sample. Our goal is to investigate the statistical trends of these molecular abundances and their dependency on the local physical conditions as a first step toward understanding the chemical evolution of dark clouds. Detailed chemical modeling of each region, which constitutes an imperative step toward deriving elemental abundances (see Fig. \ref{Fig: scheme}), will be carried out in a future paper. The observations of N$_2$H$^+$ will also be presented elsewhere since, as discussed by \citet{Fuente2019}, they require a more sophisticated modeling. We do not include H$_2$CS, because it has only been detected in a small fraction of the observed positions, preventing a statistical analysis.

\section{Sample selection}
Our project focuses on the nearby star forming regions Taurus, Perseus, and  Orion. These molecular cloud complexes were observed with Herschel and SCUBA as part of the  Gould Belt Survey \citep{Andre2010}, and accurate visual extinction (A$_V$) and dust temperature (T$_d$) maps are available \citep{Malinen2012, Hatchell2005, Lombardi2014, Zari2016}. The angular resolution of the A$_V$-T$_d$ maps ($\sim$36$''$) is  similar to that provided by the 30m telescope at 3mm allowing direct comparison of continuum and spectroscopic data. Throughout this paper we adopt A$_V\approx$N(H$_2$)$\times$10$^{-21}$ mag \citep{Bohlin1978}.

These regions are characterized as having different star formation activity and therefore different illumination. This allows us to investigate the influence of UV radiation on the gas composition. Our strategy is to observe several starless cores within each filament. By comparing the cores in the same filament, we will be able to investigate the effect of time evolution on the chemistry of dark cores (see, e.g., \citealp{Frau2012}). By comparing cores in different regions, we explore the effect of the environment on the chemistry therein. The list of selected cores is shown in Table~\ref{Table: GEMS sample}.

Within the pilot project, we observed three cuts in TMC1 and one in Barnard 1b whose data have been partially published in \citet{Fuente2019} and \citet{Navarro2020}. The data from these cuts are included in this paper for completeness. 

In total, our project includes observations towards 305 positions distributed in 27 cuts roughly perpendicular to the selected filaments. The cuts are designed to intersect the filament along one of the selected starless cores, avoiding the position of known protostars, HII regions, and bipolar outflows. They cover visual extinctions from A$_V\sim$3 mag up to A$_V\sim$200 mag in the line of sight towards the giant molecular cloud Orion A. The separation between one position and another in a given cut is selected to sample the visual extinction range in regular intervals of A$_V$. However, this is not always possible, specially for A$_V>$10 mag, where the surface density gradient is steeper. At low visual extinctions, not all the lines are detected. In this paper, we only consider 244 positions for which we are able to determine the gas density from the CS and its isotopolog observations (see Sect.~4).
In the following, we describe the observed positions in more detail:

\begin{itemize}
\item {\bf Taurus: TMC~1,  B213/L1495.} The Taurus molecular cloud (TMC), at a distance of 145 pc \citep{Yan2019}, is considered an archetypal low-mass star-forming region.  It has been the target of several cloud evolution and star formation studies \citep{Unge1987, Mizuno1995, Goldsmith2008}, being extensively mapped in CO \citep{Cernicharo1987, Onishi1996, Narayanan2008} and visual extinction \citep{Cam1999, Padoan2002, Schmalzl2010}. 

In the pilot project we centered on the filament TMC~1, which has been the target of numerous chemical studies. In particular, the positions TMC~1-CP and TMC~1-NH3  (the cyanopolyynes and ammonia emission peaks) are generally adopted as templates to compare with chemical codes \citep[e.g.,][]{Feher2016, Gratier2016, Agundez2013}. Less studied from a chemical point of view, TMC~1-C has been identified as an accreting starless core \citep{Schnee2007, Schnee2010}. Within GEMS, we observed three cuts across the TMC1-CP, TMC1-NH3, and TMC1-C (see Fig.~\ref{Fig: TMC1}). \citet{Fuente2019} carried out a complete analysis of these data to derive the gas ionization degree and elemental abundances.
 
In this work, we use  the H$_2$ column density and dust temperature maps of TMC~1 created following the process described  in \citet{Kirk2013} and \citet{Fuente2019} on the basis of Herschel \citep{Poglitsch2010, Griffin2010} data taken as part of the Herschel Gould Belt Survey \citep{Andre2010} and Planck data (c.f. \citealp{Bernard2010}). The data were convolved to the resolution of the longest wavelength, 500~$\mu$m (36 arcsec). The typical uncertainty on the fitted dust temperature was 0.3-0.4 K. The uncertainty on the column density was typically 10\% and reflects the assumed calibration error of the Herschel maps.
 
B213/L1495 is a prominent filament in Taurus that has received a lot of recent observational attention \citep{Palmeirim2013, Hacar2013, Marsh2014,Tafalla2015, Bracco2017, Shimajiry2019}. A population of dense cores previously studied in emission lines of high-dipole-moment species, such as  NH$_3$, H$^{13}$CO$^+$, and N$_2$H$^+$, are embedded in this filament \citep{Benson1989, Onishi2002, Tatematsu2004, Hacar2013, Punanova2018}. Some of these dense cores are starless, while others are associated with young stellar objects (YSOs) of different ages. This population of YSOs has been the subject of a number of dedicated studies, most recently by \citet{Luhman2009} and \citet{Rebull2010}, and by the dedicated outflow search by \citet{Davis2010}. Interestingly, the density of stars decreases from north to south suggesting a different dynamical and chemical age along the filament. The morphology of the map with striations perpendicular to the filament suggests that the filament is accreting material from its surroundings \citep{Goldsmith2008, Palmeirim2013}. \citet{Shimajiri2019} proposed that this active star-forming filament was initially formed by large-scale compression of HI gas and is now growing in mass due to the gravitational accretion of molecular gas from 
ambient cloud. We observed nine cuts along clumps \#1, \# 2,  \#5,  \#6,  \#7,  \#9,  \#12,  \#16, and  \#17 (core numbers from the catalog of \citealp{Hacar2013}). 

In this work, we use the H$_2$ column density and dust temperature maps of B213 (see Fig.~\ref{Fig: B213}) obtained by \citet{Palmeirim2013} on the basis of the Herschel Gould Belt Survey \citep{Andre2010} and Planck data (c.f. \citealp{Bernard2010}) at an angular resolution of 18.2$"$. 

\begin{figure}
%\special{psfile=14905f4.eps hoffset=-20 voffset=-230 hscale=40 vscale=35 angle=-0}
%\vspace{8.0cm}
\centering
\includegraphics[angle=0,scale=.55]{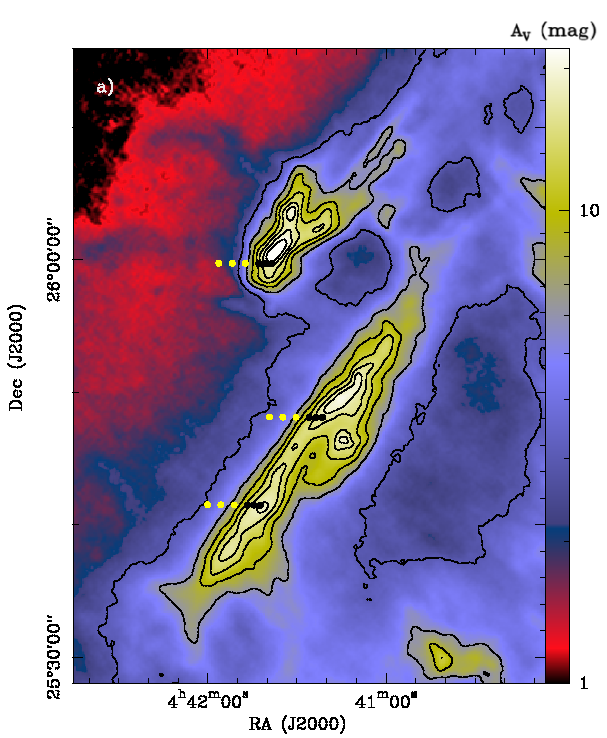}
\caption{TMC1 visual extinction map (Kirk et al., in prep). Positions observed with the 30m telescope are indicated with circles. Black circles mark positions observed only with the 30m telescope, while yellow circles indicate positions also observed with the Yebes 40m telescope. }
\label{Fig: TMC1}
%\vspace{-0.1cm}
\end{figure}

\begin{figure*}
%\special{psfile=14905f4.eps hoffset=-20 voffset=-230 hscale=40 vscale=35 angle=-0}
%\vspace{8.0cm}
\centering
\includegraphics[angle=0,scale=.29]{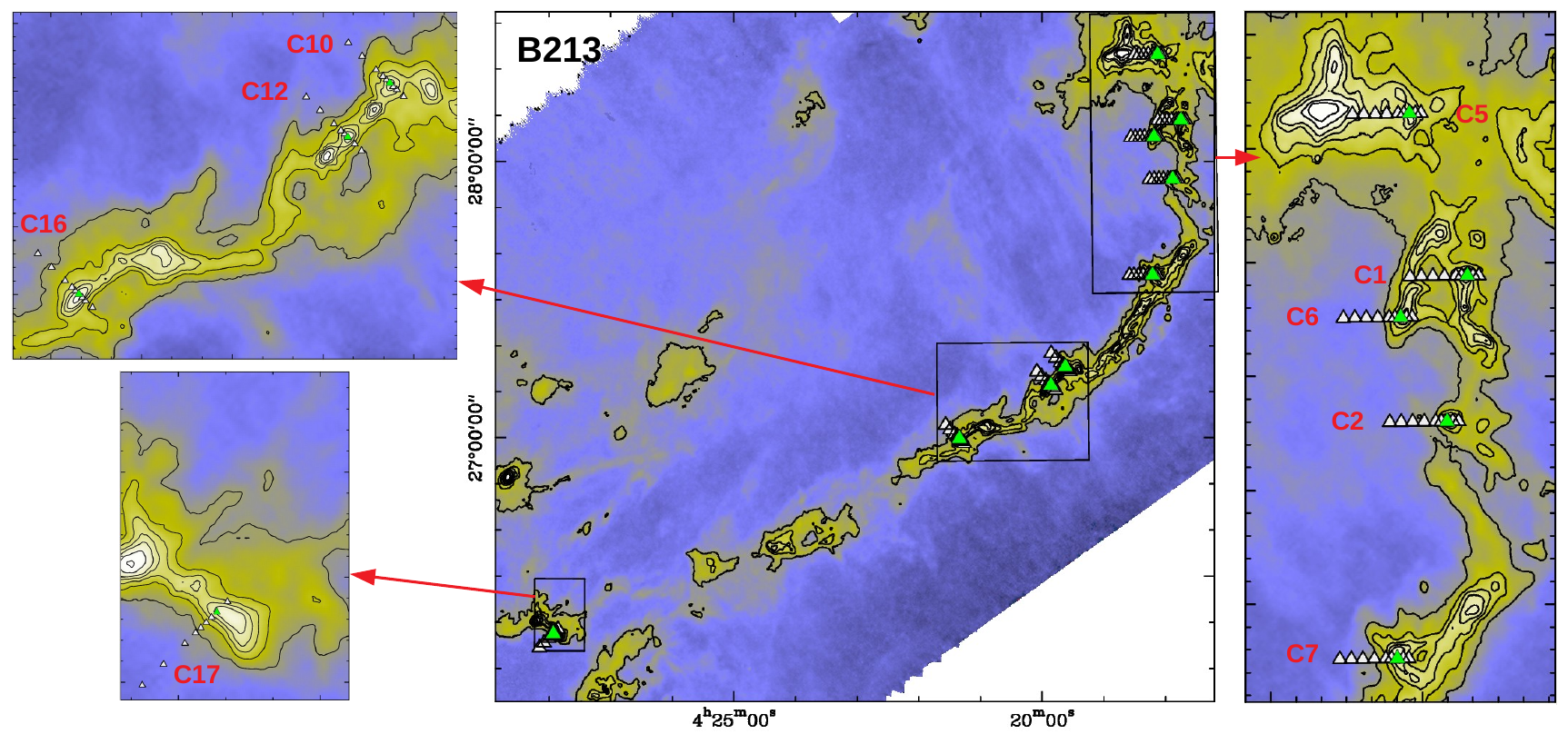}
\caption{B213 molecular hydrogen column density maps as derived by \citet{Palmeirim2013}, reconstructed at an angular resolution of 18.2$"$. General view of the region is represented at the center, and main regions of interest are enlarged. Contours are (3, 6, 9,12, 15, 20, and 25) $\times$10$^{21}$ cm$^{-2}$. Positions observed by GEMS with the 30m telescope are indicated with triangles. Green triangles represent the position of the starless cores. Labels in red indicate the cut IDs. See Table \ref{Table: GEMS sample} for further details.}
\label{Fig: B213}
%\vspace{-0.1cm}
\end{figure*}

\vskip 0.15cm

\item {\bf Perseus: Barnard 1, NGC 1333, IC348, L1448, B5.} 
The Perseus molecular cloud is a well-known star-forming cloud in the Galaxy, at a distance of 310 pc \citep{Ortiz-Leon2018}. The cloud was extensively studied using molecular line emission \citep{Warin1996, Ridge2006, Curtis2011, Pineda2008, Pineda2010, Pineda2015, Friesen2017, Hacar2017b}, star count extinction \citep{Bachiller1984}, and dust continuum emission \citep{Hatchell2005, Kirk2006, Enoch2006, Zari2016}. The molecular cloud complex is associated with two clusters containing pre-main-sequence stars: IC 348, with an estimated age of 2 Myr \citep{Luhman2003}; NGC 1333, which is younger than 1 Myr in age \citep{Lada1996, Wilking2004}; and the Per 0B2 association, which contains a B0.5 star \citep{Steenbrugge2003}. 

Perseus is the prototype low- and intermediate-mass star- forming region. The molecular cloud itself contains numerous protostars and dense cores. \citet{Hatchell2005} presented a survey of dense cores in the Perseus molecular cloud using continuum maps at 850 and 450 $\mu$m with SCUBA at the JCMT. They detected a total of 91 protostars and starless cores \citep{Hatchell2007a}. Later, \citet{Hatchell2007b} surveyed the outflow activity in the region to characterize the populations of protostars. In contrast with Taurus, a significant fraction of these protostars are associated in protoclusters, unveiling a different star formation regime. Within GEMS, we observed 11 cuts along starless cores distributed in Barnard 1, IC348, L1448, NGC 1333, and B5 (see Fig.~\ref{Fig: Perseus}). The group of cores in IC348 and NGC 1333 are close to the clusters and therefore immersed in a harsh environment, while Barnard 1 and L1448 are located in a quiescent region. 

We use the dust opacity and dust-temperature maps reported by \citet{Zari2016} in our analysis (see Fig. \ref{Fig: Perseus}). In order to derive the molecular hydrogen column density from the dust opacity at 850$\mu$m ($\tau_{850}$), we used expression (7) of  \citet{Zari2016} and A$_V$=A$_K$/0.112. This expression gives accurate values for low extinctions  (A$_V<$ 10 mag) but may underestimate their value towards the extinction peaks. In the range of values considered in Perseus, A$_V\sim$3$-$30 mag, the uncertainty in the 
values of A$_V$ is a factor of two.

\begin{figure*}
\centering
%\special{psfile=14905f4.eps hoffset=-20 voffset=-230 hscale=40 vscale=35 angle=-0}
%\vspace{8.0cm}
\includegraphics[angle=0,scale=.3]{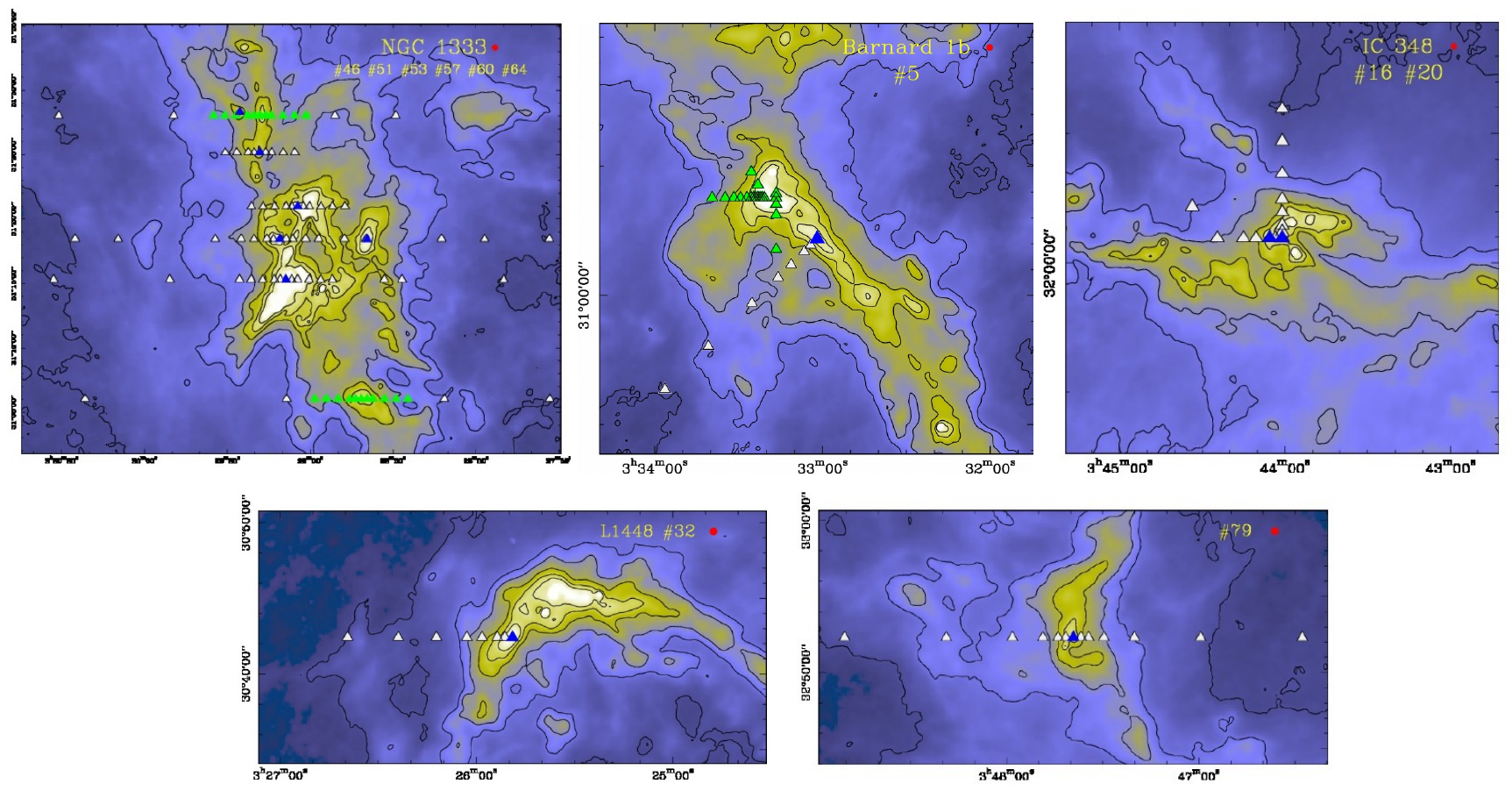}
\caption{Perseus filaments (from left to right and top to bottom: NGC\,1333, Barnard 1, IC348, L1448, and B5) dust opacity maps at 850$\mu$m by \citet{Zari2016}, convolved at an angular resolution of 36$"$. Contours are (0.056, 0.13, 0.24, 0.56, 1.01, and 1.6) $\times$10$^{-3}$, which according to expression (7) from \citet{Zari2016}  corresponds to visual extinctions of $\sim$5, 7.5, 10, 15, 20, and 25 mag, respectively. Positions observed with the 30m telescope are indicated with triangles. Blue triangles represent the positions of the starless cores. }
\label{Fig: Perseus}
%\vspace{-0.1cm}
\end{figure*}

\vskip 0.15cm
\item {\bf Orion A.} 
The Orion star-forming region is the most massive and most active star-forming complex in the local neighborhood (e.g., \citealp{Maddalena1986, Brown1995, Bally2008,Lombardi2011, Kainulainen2017, Friesen2017, Monsch2018, Getman2019, Karnath2020, Tobin2020, Hacar2020}). It contains the nearest massive star-forming cluster to Earth, the Trapezium cluster (e.g., \citealp{Hillenbrand1997, Lada2000,  Muench2002, Dario2012, Robberto2013, Zari2019}), at a distance of 388 pc \citep{Kounkel2017}. Traditionally, the “Orion nebula” refers to the visible part of the region, the HII region, powered by the ionizing radiation of the Trapezium OB association. It is part of a much larger complex, referred to as the Orion molecular cloud (OMC), itself formed by two giant molecular clouds: Orion A hosting the Orion nebula and the more quiescent Orion B (see, e.g., \citealp{Pety2017}), both lying at the border of the Eridanis super bubble \citep[see e.g.,][]{Bally2008,Ochsendorf2015,Pon2016}.

Orion A has been extensively mapped in molecular lines using single-dish telescopes \citep{Hacar2017a, Goicoechea2019, Nakamura2019, Tanabe2019, Ishii2019, Hacar2020} and large millimeter arrays \citep{Kirk2017, Hacar2018,Monsch2018,Suri2019, Kong2019}. Different clouds have been identified within Orion A based on  millimeter, submillimeter, and infrared observations. Orion molecular cloud 1 (OMC~1) was identified as a dense gas directly associated with Orion KL \citep{Wilson1970, Zuckerman1973, Liszt1974}, then OMC~2 \citep{Gatley1974} and OMC~3 \citep{Kutner1976} were detected as subsequent clumps in CO emission located about 15$'$ and 25$'$ to the north of OMC~1. The $^{13}$CO (J =1$\rightarrow$0) observations by \citet{Bally1987} revealed that these clouds consist of the integral-shaped filament (ISF) of molecular gas,  itself part of a larger filamentary structure extending from north to south over 4$^\circ$. After that, the SCUBA maps at 450 $\mu$m and 850 $\mu$m presented concentrations of submillimeter continuum emission in the southern part of the integral-shaped filament, which are now referred to as OMC~4 \citep{Johnstone1999} and OMC~5 \citep{Johnstone2006}.

Three cuts along OMC-2 (ORI-C3), OMC-3 (ORI-C1), and OMC-4 (ORI-C2) were observed within GEMS (see Fig. \ref{Fig: Orion}). These cuts avoid the protostars and stars in this active star-forming region, probing different environments because of their different distance from the Orion nebula. In our analysis, we use the dust opacity and dust temperatures maps reported by \citet{Lombardi2014}. Dust temperatures along ORI-C3 are higher than towards ORI-C1 and ORI-C2 with
values always $>$ 30 K. The values of the molecular hydrogen column density are derived
from the dust opacity at 850$\mu$m using expression $A_{\rm K}=2640 \times \tau_{850} + 0.012$ of  \citet{Lombardi2014} and A$_V$=A$_K$/0.112.

\begin{figure}
%\special{psfile=14905f4.eps hoffset=-20 voffset=-230 hscale=40 vscale=35 angle=-0}
%\vspace{8.0cm}
\centering
\includegraphics[angle=0,scale=.33]{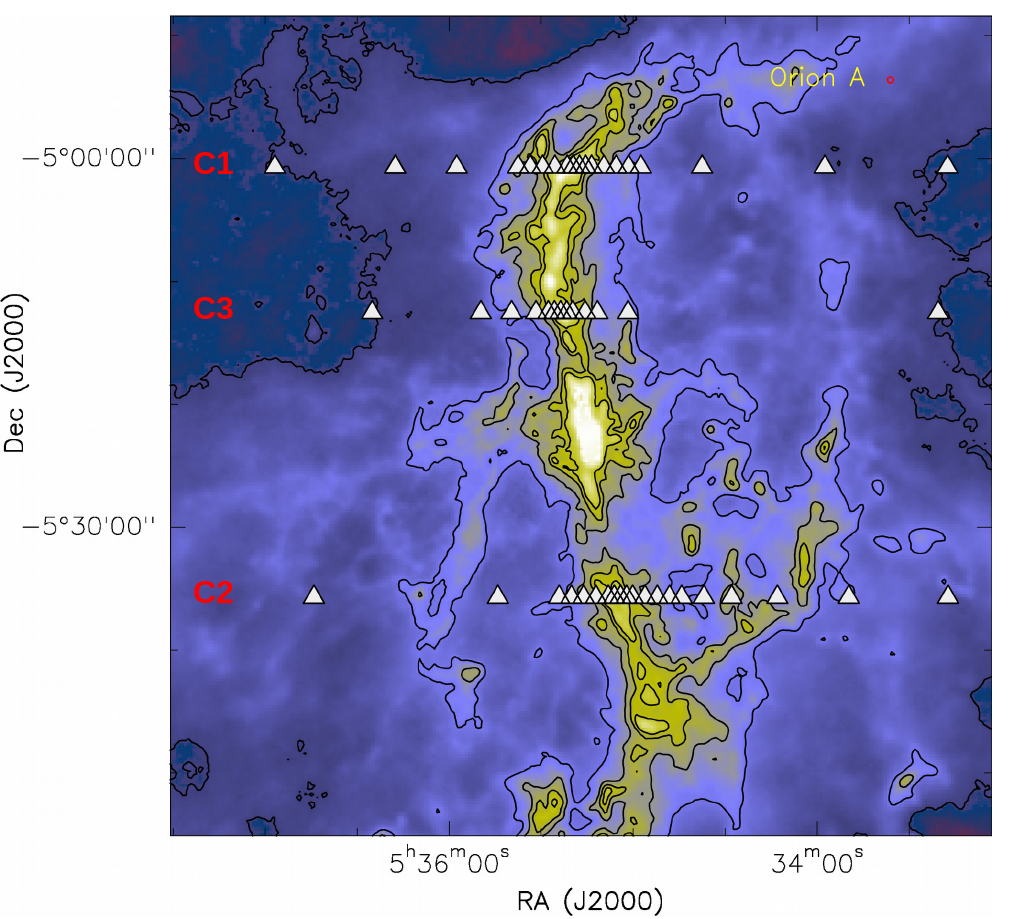}
\caption{Orion dust opacity map at 850$\mu$m by \citet{Lombardi2014}, convolved at an angular resolution of 36". Contours are (0.056, 0.24, 0.56, 1.36, and 1.61) $\times$10$^{-3}$, which according to \citet{Lombardi2014} correspond to visual extinctions of $\sim$1.3, 5.6, 13.2, 23.8, and 38 mag. Positions observed with the 30m telescope are indicated with triangles. Labels in red indicate the cut IDs. }
\label{Fig: Orion}
%\vspace{-0.1cm}
\end{figure}

\end{itemize}

\begin{table*}
\caption{Cores included in the GEMS sample and observation cuts associated to them and shown in Fig.~\ref{Fig: TMC1}, Fig.~\ref{Fig: B213}, Fig.~\ref{Fig: Perseus},
and Fig.~\ref{Fig: Orion}. N$_{\rm o}$ indicates the total number of points observed in 
the corresponding cut. N$_{\rm v}$ indicates the number of points where the molecular hydrogen density could be derived (see Sect. \ref{Sec.Physical conditions}).}
\label{Table: GEMS sample}
\centering
\begin{tabular}{llrlllll}\\
\hline\hline
\noalign{\smallskip}
Cloud & Core & \multicolumn{2}{c}{Coordinates} & Other names & Cut & N$_{\rm o}$ & N$_{\rm v}$ \\
& ID & RA (J2000) & Dec (J2000) & & & & \\
\hline
\noalign{\smallskip}                          
%
%\multicolumn{6}{l}{{\bf Taurus}} \\
TMC1        &  CP   & 04:41:41.90   &  $+$25:41:27.1    & & C1          &    6& 6 \\
            &  NH3  &  04:41:21.30   &  $+$25:48:07.0    & &  C2           &   6& 6  \\   
            &  C    &  04:41:38.80   &  $+$25:59:42.0   & &   C3          &  6 & 6  \\
B\,213$^1$  &  \#1  &  04:17:41.80   &     $+$28:08:47.0     & 5   &  C1    &   9 & 9  \\   
            &  \#2  &   04:17:50.60   &     $+$27:56:01.0     &  $-$ &  C2   &   9 & 7  \\
            &  \#5  &  04:18:03.80   &     $+$28:23:06.0     & 7  &   C5   &  9 & 9  \\
            &  \#6  &  04:18:08.40   &     $+$28:05:12.0     &  8  &  C6   &   9 &  5  \\    
            &  \#7  &  04:18:11.50   &     $+$27:35:15.0     & 9  &  C7     &   9 & 8  \\   
            &  \#10 &  04:19:37.60   &     $+$27:15:31.0     & 13a &  C10    &   9 &  8  \\  
            &  \#12 &  04:19:51.70   &     $+$27:11:33.0     & $-$ &   C12    &  9 & 6  \\  
            &  \#16 &  04:21:21.00   &     $+$27:00:09.0     & $-$ &  C16    &  9  &  9  \\
            &  \#17 &   04:27:54.00  &    $+$26:17:50.0      & 26b &   C17    &  9 & 5  \\
%\multicolumn{6}{l}{{\bf Perseus}$^3$} \\                   
L1448$^2$   &  \#32 &  03:25:49.00   &     $+$30:42:24.6      &   &    C1   & 8 & 7 \\
NGC\,1333$^2$   &       &   03:29:18.26   &      $+$31:28:02.0     &  &   C1    &   21 & 18 \\    
            &       &  03:28:41.60   &      $+$31:06:02.0     &   &   C2  &  17 & 13 \\
            &  \#46 &  03:29:11.00   &   $+$31:18:27.4     &   SK20$^3$  &   \multirow{2}{*}{C3}  & \multirow{2}{*}{17} & \multirow{2}{*}{11} \\
            &  \#60 & 03:28:39.40    &      $+$31:18:27.4     &   &      & & \\%& \\%&\\    
            &  \#51 &  03:29:08.80   &      $+$31:15:18.1     &   SK16    &    C4  & 16  & 13  \\
            &  \#53 & 03:29:04.50  &      $+$31:20:59.1     &   SK26   &   C5  & 11 &  11  \\  
            &  \#57 &  03:29:18.20     &      $+$31:25:10.8     &    SK33  &   C6  & 9  & 9  \\ 
            &  \#64 &  03:29:25.50     &      $+$31:28:18.1     &  &   C7      & 1 & 1  \\ 
Barnard 1$^2$   &  1b   &  03:33:20.80 & $+$31:07:34.0  & &  C1   & 18  & 18 \\ 
            &       &  03:33:01.90     &    $+$31:04:23.2    &  &    C2      & 8   & 6  \\
Barnard 5$^2$   &  \#79 &  03:47:38.99    &      $+$32:52:15.0    &   &    C1     &  13 &  9 \\ 
IC\,348$^2$     &  \#1  &    03:44:01.00    &   $+$32:01:54.8   &  &  \multirow{2}{*}{C1}  & \multirow{2}{*}{14} & \multirow{2}{*}{11} \\
            &  \#10 &  03:44:05.74   &       $+$32:01:53.5      &    &     &   &  \\                   
Orion A     &       &   05:35:19.54   &      $-$05:00:41.5        & &   C1         & 20 & 10  \\
            &       &  05:35:08.15    &      $-$05:35:41.5        & &   C2         & 20 & 14  \\
            &       & 05:35:23.68  &     $-$05:12:31.8     &  &    C3         & 13 & 9  \\

\hline
\noalign{\smallskip}
TOTAL       &              &    & &  &  & 305 & 244 \\                  
\hline \hline
\end{tabular}

$^1$B213 core IDs are from \citet{Hacar2013}. IDs indicated in "Other names" column are from \citet{Onishi2002}. \\
\noindent
$^2$Perseus core IDs (L 1448, NGC 1333, Barnard 1, Barnard 5, IC348) are from \citet{Hatchell2007a}. \\
\noindent
$^3$NGC\,1333 core IDs indicated in "Other names" column are from \citet{Sandell2001}.
\end{table*}

\section{Observations}

The 3\,mm and 2\,mm observations were carried out using the IRAM 30-m telescope at Pico Veleta (Spain) during three observing periods in July 2017, August 2017, and February 2018. The observing mode was frequency switching with a frequency throw of 6 MHz well adapted to removing standing waves between the secondary mirror and the receivers. The Eight MIxer Receivers (EMIR) and the Fast Fourier Transform Spectrometers (FTS) with a spectral resolution of 49~kHz were used for these observations. The intensity scale is T$_{MB}$, which is related to T$_A^*$ by $T_{MB} =(F_{eff} / B_{eff})T_A^*$ (see Table B.1 in \citealp{Fuente2019}). The difference between T$_A^*$  and  T$_{MB}$ is $\approx$ 17 \% at 86 GHz and 27\% at 145 GHz. The uncertainty in the source size is included in the line intensity errors, which are assumed to be $\sim$20\%. 

For TMC~1 and Barnard 1, we use observations of the CS 1$\rightarrow$0 line carried out with the Yebes 40m radiotelescope \citep{Tercero2020}. The 40m telescope is equipped with HEMT receivers for the 2.2-50 GHz range, and a superconductor-insulator-superconductor (SIS) receiver for the 85-116 GHz range. Single-dish observations in K-band (21-25 GHz) and Q-band (41-50 GHz) can be performed simultaneously. The backends consisted of a FFTS covering a bandwidth of $\sim$2 GHz in band K and $\sim$9 GHz in band Q, with a spectral resolution of $\sim$38 kHz. Detailed information about the setups observed in the IRAM 30m and Yebes 40m telescopes and the telescopes parameters were presented in \citet{Fuente2019}. 

\section{Physical conditions: molecular hydrogen density}
\label{Sec.Physical conditions}

In order to derive the gas physical conditions, we use the line intensities of the observed CS, C$^{34}$S, and $^{13}$CS lines. CS has been widely used as a density and column density tracer in the interstellar medium \citep{Linke1980, Zhou1989, Tatematsu1993, Zinchenko1995, Anglada1996, Bronfman1996, Laundhardt1998, Shirley2003, Bayet2009, Wu2010, Zhang2014, Scourfield2020}. We fit the lines using the molecular excitation and radiative transfer code RADEX \citep{Tak2007}, which do not consider local thermodynamic equilibrium (LTE), and the collisional coefficients calculated by \citet{Denis2018}. During the fitting process, we fix the ratios  $^{12}$C/$^{13}$C=60 and $^{32}$S/$^{34}$S= 22.5 \citep{Savage2002, Gratier2016}. Isotopic fractionation is not expected to be important for sulfur. The CS molecule is enriched in $^{34}$S at early time because of the $^{34}$S$^+$ + CS reaction but not at the characteristic times of dense clouds, which justifies the use of a fixed C$^{34}$S/CS ratio to derive molecular hydrogen densities and abundances \citep{Loison2019b}. More controversial is the $^{12}$CS/$^{13}$CS ratio; the adopted value is consistent with the results of \citet{Gratier2016} in TMC~1 and \citet{Agundez2019} in L~483. This value is also consistent with chemical predictions for typical conditions in dark clouds and evolutionary times of approximately a few hundred thousand years \citep{Colzi2020, Loison2020}. We assume a beam-filling factor of 1 for all transitions (the emission is more extended than the beam size). 

\begin{figure*}
%\special{psfile=14905f4.eps hoffset=-20 voffset=-230 hscale=40 vscale=35 angle=-0}
%\vspace{8.0cm}
\centering
\includegraphics[scale=.9]{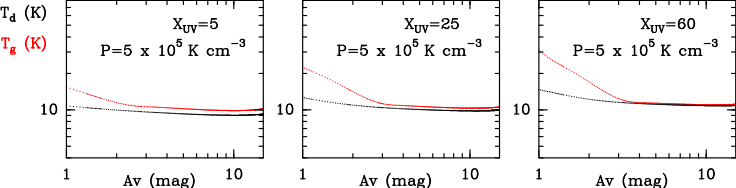}
\caption{Dust and gas temperature derived using the Meudon PDR 1.5.4 code for a plane-parallel isobaric cloud illuminated from only one side.  The plotted dust temperature is a weighted average of the dust temperatures calculated for different grain sizes, assuming a  grain size distribution,  $n_{gr} \propto a^{-3.5}$ \citep{Mathis1977}, where $n_{gr}$ is the number of grains with size $a$, and the minimum  and maximum grain sizes are given by $a$ =10$^{-7}$cm and 3$\times$10$^{-5}$~cm, respectively. The incident UV flux and thermal pressure of each calculation are indicated in the panels. The  values have been selected to represent the physical conditions in Taurus ($\chi_{\rm UV} \sim$5), Perseus ($\chi_{\rm UV} \sim$25), and Orion ($\chi_{\rm UV} \sim$50), where $\chi_{\rm UV}$  is the incident UV field in units of the Draine field \citep{Draine1978}. }
\label{grains}
%\vspace{-0.1cm}
\end{figure*}

In addition, we assume that gas and dust are thermalized, that is, that the  kinetic temperature T$_k$ is equal to the dust temperature derived from far-infrared and millimeter observations (TMC~1: \citealp{Fuente2019}; B~213: \citealp{Palmeirim2013}; Perseus: \citealp{Zari2016}; Orion: \citealp{Lombardi2014}). This assumption might underestimate the gas temperature at the low extinctions  where the gas can be heated by the photoelectric effect to temperatures higher than the dust temperature (see, e.g., \citealp{Okada2013}). In order to test the reliability of this assumption we carried out thermal-balance calculations for three representative cases. Figure~\ref{grains} shows the dust and the gas temperatures calculated using the Meudon PDR (1.5.4) \citep{LePetit2006,Goicoechea2007,GonzalezGarcia2008, Bourlot2012}. The three panels show the output for three models that are representative of the physical conditions in Taurus \citep{Fuente2019, Navarro2020}, Perseus \citep{Navarro2020}, and Orion. For Orion, we adopt $\chi_{UV} \sim 60$, which is the incident Draine field estimated in the Horsehead nebula \citep{Pety2005,Goicoechea2006,Goicoechea2009a,Goicoechea2009b,Guzman2011,Guzman2012a,Guzman2012b,Guzman2013,Legal2017,Riviere2019}. In every case the value of the cosmic ray ionization rate is set to $\zeta$(H$_2$) = 5$\times$10$^{-17}$ s$^{-1}$. For all cases it could be observed that T$_g$ $\sim$T$_d$ within $\sim$1~K when considering the region with A$_V >$ 4~mag. It should be noted that these calculations consider a plane-parallel slab illuminated from one side. In the more realistic case of a cloud illuminated from the front and back, the visual extinction at which  T$_g$ $\sim$T$_d$  would be $\sim$ 8~mag. In the outer part of the cloud, the gas temperature is higher than the dust temperature by an amount that varies with the incident UV field. Following our calculations,  for A$_V$ $\sim$ 2~mag,   T$_g$ $\sim$T$_d$ + 5~K in Taurus,  T$_g$ $\sim$T$_d$ + 10~K in Perseo, and T$_g$ $\sim$T$_d$ + 15~K   in Orion. Most of our points are located in the region $>$ 8~mag. Moreover, the lines of H$^{13}$CO$^+$, HC$^{18}$O$^+$, H$^{13}$CN, HCS$^+$, and OCS are only detected towards A$_V>$ 8~mag. Therefore, we consider that our assumption is reasonable for the goals of this paper, which is focused on statistical trends.

\begin{figure}
%\special{psfile=14905f4.eps hoffset=-20 voffset=-230 hscale=40 vscale=35 angle=-0}
%\vspace{8.0cm}
\includegraphics[scale=.38]{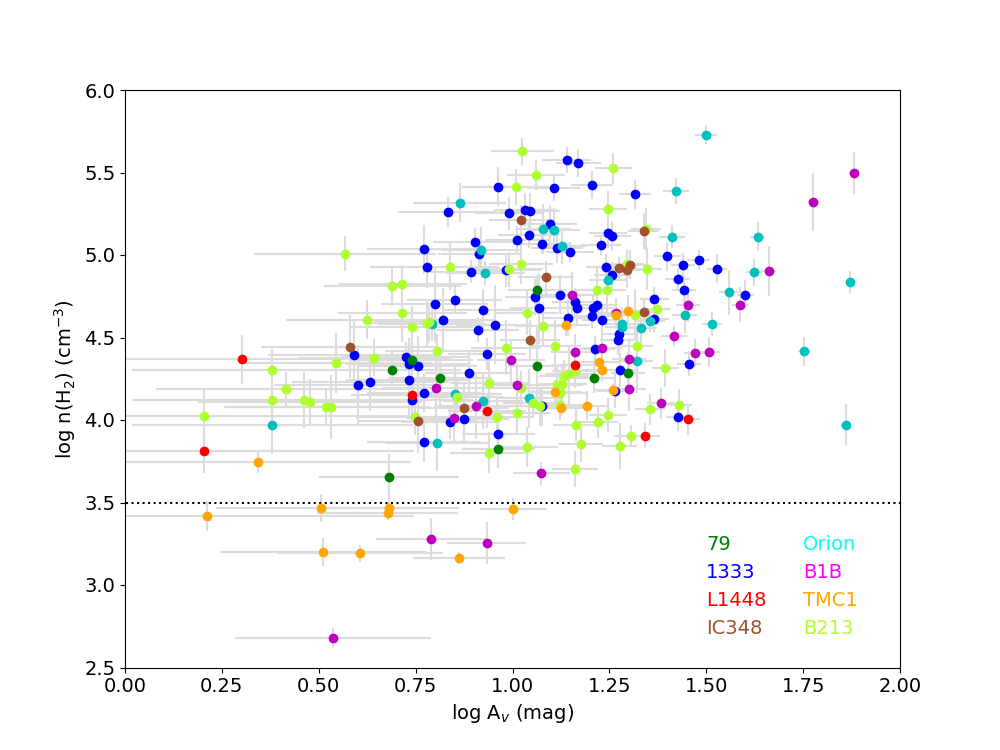}
\caption{Relation between derived molecular hydrogen densities and visual extinctions for the sample used in this study. Colors indicate the different observed regions as shown in the plot legend.  Densities < 3$\times$10$^3$ cm$^{-3}$ are only measured in TMC 1 and Barnard 1b as a consequence of our methodology and the limitations of our dataset, which is incomplete for the rest of the regions (see text). This value is marked with a dotted horizontal line.}
\label{Densities_extinctions}
%\vspace{-0.1cm}
\end{figure}

In our calculations, we let n(H$_2$) and  N(CS) vary as free parameters and explore their parameter space following the Monte Carlo Markov Chain (MCMC) methodology with a Bayesian inference approach. In particular, we used the emcee \citep{Foreman2012} implementation of the Invariant MCMC Ensemble sampler methods by \citet{Goodman2010}. This method was already used in \citet{Riviere2019} and \citet{Navarro2020}, and allowed us to estimate the density as long as the two transitions of CS, J=2$\rightarrow$1, and 3$\rightarrow$2 are detected, which reduced the total number of valid spatial points to 244. In TMC 1 and Barnard 1b, we complemented the J=2$\rightarrow$1 and 3$\rightarrow$2 observations with data of the J=1$\rightarrow$0 line as observed with the 40m Yebes telescope. In these two clouds, we were able to estimate densities at the edge of the cloud down to A$_V = 3$ mag, where the J=3$\rightarrow$2 line of CS is not detected. These estimates were published in \citet{Fuente2019} and \citet{Navarro2020}, and are included in the present analysis. For the remaining sources, we do not have CS J=1$\rightarrow$0 data, and we are only sensitive to densities of n(H$_2$)$\sim$10$^4$ cm$^{-3}$ or larger. In Fig.~\ref{Densities_extinctions} we plot the derived molecular hydrogen densities as a function of the visual extinction, with colours indicating the different observed filaments. As mentioned above, densities $<$ 3$\times$10$^3$ cm$^{-3}$ are only measured in TMC 1 and Barnard 1b as a consequence of our methodology and the limitations of our dataset. There is a clear trend of increasing hydrogen density with visual extinction, with higher densities in the inner layers of the clouds. However, the dispersion is large, with values of the molecular hydrogen density varying by a factor of $>$ 30 for a given visual extinction. This dispersion remains even if only the points of a given cloud are considered, which suggests that it is not the result of mixing clouds in different environments but the consequence of a complex density structure with different gas layers along the line of sight.

\begin{figure*}
\includegraphics[scale=.36]{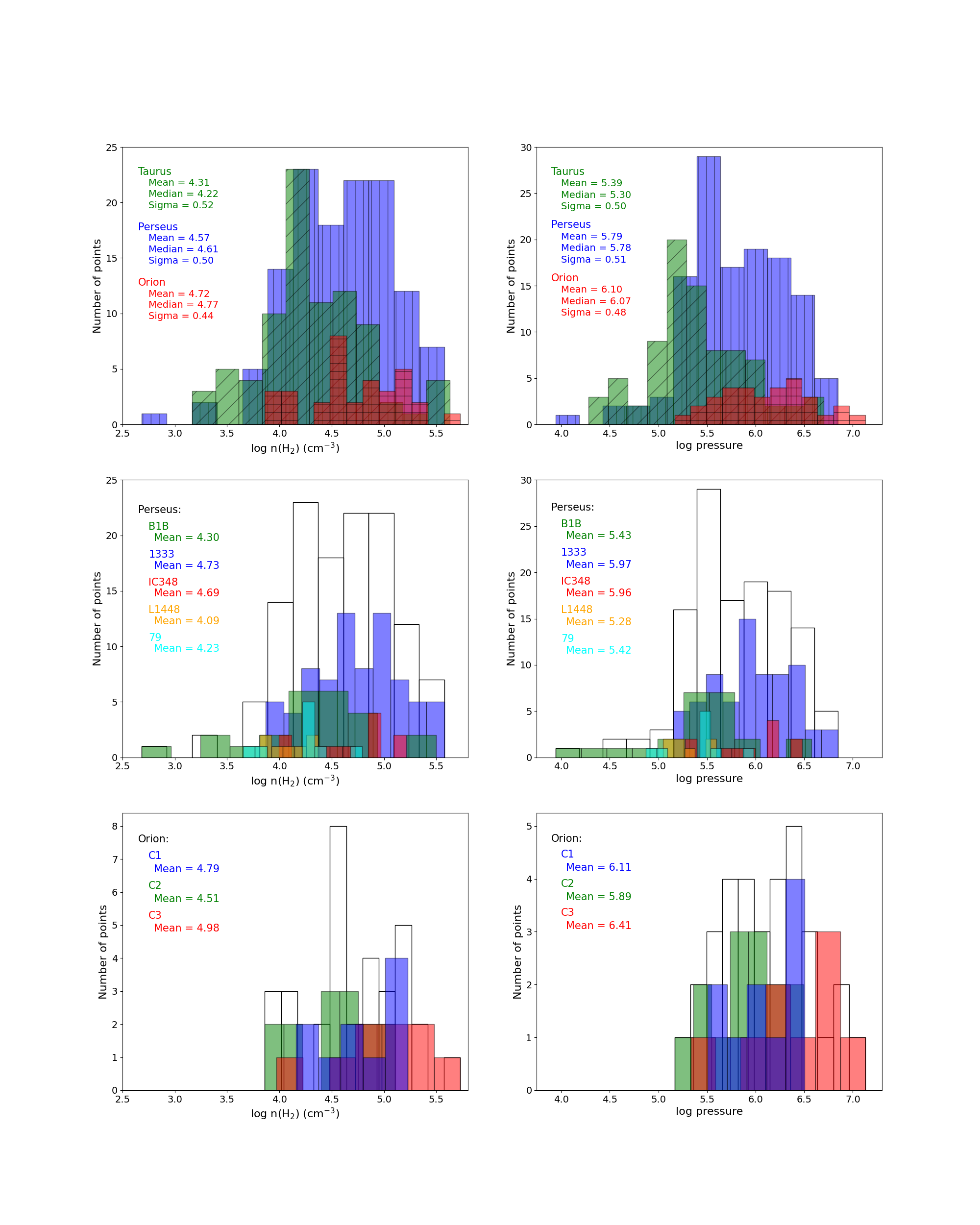}
\caption{Histograms of the derived molecular hydrogen densities (left column) and corresponding pressures (right column). Histograms comprise values for the whole sample indicating the corresponding molecular complexes (top row), clouds in Perseus (middle row), and cuts of the Orion cloud (bottom row). In the middle and bottom rows, white histograms show the distribution for the molecular cloud complex to facilitate the comparison. Statistical parameters of the different distributions are indicated in each histogram legend.}
\label{Fig: Histograms_dens_pres}
%\vspace{-0.1cm}
\end{figure*}

The distribution of the molecular hydrogen densities and gas thermal pressures ($n(H_2) \times T_{\rm K}$) obtained for the 244 points is represented in the top row of Fig.~\ref{Fig: Histograms_dens_pres}, color-coded according to the molecular cloud that they belong to, with legends indicating the mean, median, and standard deviation values of their corresponding distributions in logarithmic units. The densities in Taurus show a peaky distribution with a low mean density value, as expected for low-mass star-forming regions. Perseus has higher values of densities and pressures, with a wider distribution. The highest values of density and pressure and the most flattened distribution is observed in Orion. In order to explore the origin of the wide distribution in Perseus, we made the density and pressure histograms differentiating the individual clouds of this region (middle row in Fig.~\ref{Fig: Histograms_dens_pres}). NGC\,1333 and IC\,348, with higher temperatures and extinctions, have density and pressure values closer to those of Orion, while low temperature regions such as Barnard\,1b, L\,1448, and Barnard\,5 show values more similar to those obtained in Taurus. The wide distribution shown by Perseus therefore seems to be produced by the different contributions of its five observed regions, comprising a certain range of physical parameters. We carried out the same analysis in Orion, although we only observed three cuts without statistical significance. The three peaks observed in the pressure distribution of Orion (bottom row of Fig.~\ref{Fig: Histograms_dens_pres}) 
indeed correspond to the three observed cuts. As one would expect, the cut located closer to the Orion nebula (cut 3) is the one with the highest pressure. The cut with the lowest pressure, that is, the most similar to low-mass star forming regions, is the one in OMC~4 (cut 2). The cut in OMC~3 presents intermediate conditions. This plot suggests that the range of physical conditions in high-mass and intermediate-mass star forming regions is wider than the one observed in the low-mass star forming regions, which is surely related to the feedback of the recently formed intermediate- and high-mass stars in the environment.

\section{Molecular column densities}
\label{Sec: Molecular column densities}

The fitting of the CS (and its isotopologs C$^{34}$S and $^{13}$CS) lines described in Sect. \ref{Sec.Physical conditions} allows us to accurately determine the CS column density values, and therefore its molecular abundances, for each of the 244 points where the MCMC method can be applied. Considering the derived molecular hydrogen densities, we can also determine the molecular column densities and the molecular abundances for a set of species for which only one line is observed: $^{13}$CO, C$^{18}$O, HCO$^+$, H$^{13}$CO$^+$, HC$^{18}$O$^+$, H$^{13}$CN, HNC, HCS$^+$, SO, $^{34}$SO, H$_2$S, and OCS. We use the RADEX code and the collisional coefficients included in Table \ref{Table:collisional coefficients}. Uncertainties are calculated taking into account the errors in the measurement of the integrated line intensities as well as the systematic errors of 10\%-20\% due to the flux calibration. 

In the following, we investigate the relationship between the obtained molecular abundances and the molecular cloud physical parameters: kinetic temperature, extinction, and molecular hydrogen density, considering each molecular cloud separately and the statistical trends observed for the complete dataset. As a first approach, in this section we analyze the correlations and trends that can be observed in the corresponding figures in a qualitative way. A quantitative discussion based on numerical values of correlation coefficients is provided in Sect. \ref{Sec: Statistical correlations}. It is important to note that in the particular case of the CS, we obtain the molecular abundance by fitting CS and its isotopologs C$^{34}$S and $^{13}$CS simultaneously, which implies that we are already taking into account the influence of line opacity. We also assume that  the CS, C$^{34}$S, and $^{13}$CS line emission comes from the same region. The method applied for the other species is different, as mentioned above, and this should be considered in the interpretation of the results. 

\begin{table}
\caption{References for the collisional rate coefficients used for the volume density estimates.}
\label{Table:collisional coefficients}
\centering
\begin{tabular}{lll}\\%l}\\
%\multicolumn{3}{l}{Table 1. GEMS sample} \\ 
\hline\hline
\noalign{\smallskip}
Molecule &  & Reference \\%& CD(CS)\\
\hline
\noalign{\smallskip}                          
CS & & \citet{DenisAlpizar2013}   \\
%CO & & Yang et al. (2010) \\
$^{13}$CO & & \citet{Yang2010} \\
C$^{18}$O  & & \citet{Yang2010} \\
HCO$^+$  & & \citet{Yazidi2014}\\
H$^{13}$CO$^+$  & & \citet{Yazidi2014} \\
HC$^{18}$O$^+$  & & \citet{Flower1999} \\
HCS$^+$ && \citet{Flower1999} \\
H$^{13}$CN && \citet{HernandezVera2014, HernandezVera2017} \\
SO & & \citet{Lique2007} \\
$^{34}$SO & & \citet{Lique2007} \\
HNC & & \citet{Dumouchel2011} \\
     & & \citet{ HernandezVera2014} \\
OCS & & \citet{Green1978} \\
o-H$_{2}$S & & \citet{Dagdigian2020} \\

\hline
\noalign{\smallskip}
\end{tabular}
\end{table}

\subsection{CS}
\label{Subsec: CS}

This is a diatomic molecule with well-known collisional coefficients \citep{DenisAlpizar2013} and, as already mentioned in Sect. \ref{Sec.Physical conditions}, is widely used as a density and column density tracer in the interstellar medium. In our Galaxy, CS is the most ubiquitous sulfur compound, the only one that is commonly detected in photodissociation regions \citep[PDRs;][]{Goicoechea2016, Riviere2019} and protoplanetary disks \citep{Dutrey1997, Fuente2010, Dutrey2011, Guilloteau2016, Teague2018, Phuong2018, Legal2019}. Therefore, a complete understanding of its chemistry would be of great value for estimating the physical conditions of the gas and sulfur depletion in many astrophysical environments.  In the following, we investigate the behavior of CS abundance in the GEMS sample.

\begin{figure*}
\includegraphics[scale=.46]{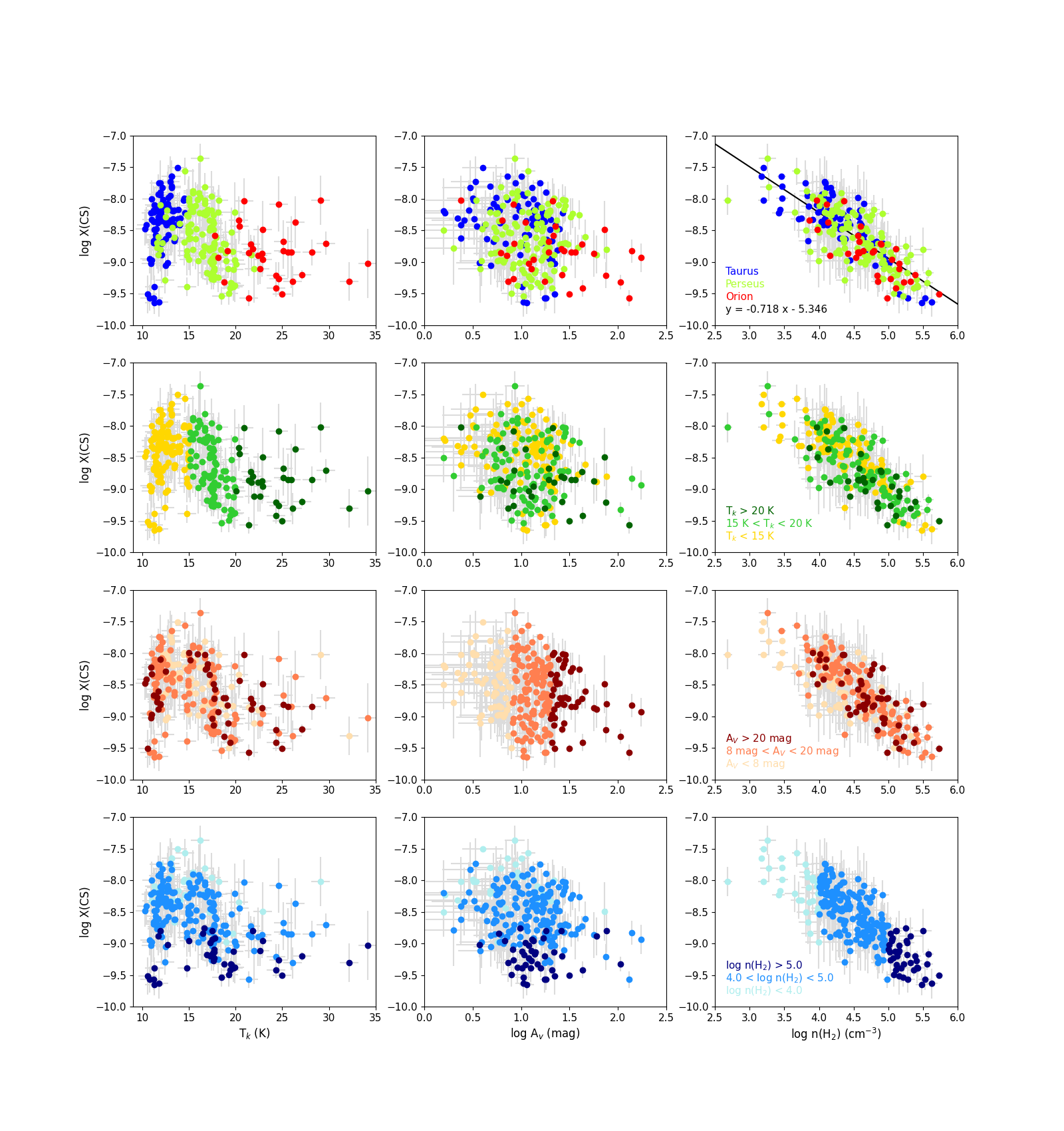}
\caption{Relation of the CS molecular abundance to cloud physical parameters: kinetic temperature (left column), extinction (middle column), and molecular hydrogen density (right column). From top to bottom, the dataset is color-coded according to the molecular cloud of the points (first row), bins of kinetic temperature (second row), bins of extinction (third row), and bins of molecular hydrogen density (last row), as indicated in the corresponding legends. The black line in the top right plot shows the linear correlation between X(CS) and n(H$_2$) for the whole sample, with the correlation parameters indicated in the legend.}
\label{Fig: Molecular_abundances_CS}
%\vspace{-0.1cm}
\end{figure*}

The relationships between the derived CS molecular abundances and the main physical parameters of the clouds (kinetic temperature, extinction, and molecular hydrogen density) are represented in Fig.~\ref{Fig: Molecular_abundances_CS}. The rows of the figure represent the same relations but classifying the data following different criteria: first, the molecular cloud to which the points belong, and then dividing the dataset considering bins of temperature, extinction, or density. These divisions allow us to study the possible influence of the variation of these parameters in the whole sample, beyond the particular environment of each cloud. We consider bins of temperature with limits of 15 K and 20 K, which approximately describe different trends observed in the plots. Bins of extinction are established below 8 magnitudes (translucent cloud), between 8 and 20 magnitudes (dense core in low-mass star forming regions), and above 20 magnitudes (dense cores in intermediate and massive star-forming regions). Bins of molecular hydrogen density are considered with limits of 4 and 5 in logarithmic scale, which describe the range of values that we observe in our data. 

The decreasing linear relation between CS molecular abundance and molecular hydrogen density is very clear. This trend, the linear fitting of which for the whole dataset is included in the top-right panel of Fig. \ref{Fig: Molecular_abundances_CS}, is common to the three molecular clouds, although each cloud presents differentiated physical conditions. Moreover, albeit with a dispersion by a factor of three, this trend remains for almost three orders of magnitude. Therefore, this parameter is the main driver of the changes in the CS abundance in our sample.

In  Fig.~\ref{Fig: Molecular_abundances_CS}, we also show X(CS) as a function of the gas kinetic temperature. A possible correlation is observed between X(CS) and T$_k$ up to $\sim$14~K. Afterwards the abundance decreases with temperature in the range T$_{k}\sim$14$-$20~K. Beyond this point, T$_k$>20~K,  the scatter increases and no clear trend is observed. Interestingly enough, the classification of points according to their molecular cloud precisely reproduces this behavior:  points with T$_k$<14 K belong to Taurus and the low-temperature regions of Perseus, L1448 and B1B. In these cuts, X(CS) is increasing with T$_k$. It should be noted that density and T$_k$ are anti-correlated in these regions, with the lowest values of T$_k$ being associated with the highest densities. This trend with T$_k$ is then related with the overall trend of X(CS) decreasing with n(H$_2$). On the other hand, most of the Perseus points show the decreasing relation between 15~K and 20~K.  We interpret this behavior as the consequence of the CS photodissociation at the illuminated cloud edge. We remind that G$_0$ is higher in star forming regions of Perseus than in Taurus \citep{Navarro2020} and in our sample of starless cores, the highest temperatures are found at the lowest visual extinctions.  Lastly, points belonging to Orion show a high scatter and no clear trend with temperature. Interestingly, the Orion points follow the general anti-correlation between X(CS) and gas density quite well, confirming density as the dominant parameter of the CS chemistry.

The depletion of CS has been widely studied in starless cores, and the X(CS)/X(N$_2$H$^+$) abundance ratio has been proposed as an evolutionary diagnostic for these objects \citep{Tafalla2002, Tafalla2004, TafallaSantiago2004, Hirota2006, Heithausen2008, Kim2020}. Our data show that the CS abundance decreases almost linearly with the density for almost three orders of magnitude. Moreover, this trend remains valid for regions regardless of the star formation activity of the region, providing a tool with which to predict density  and to subsequently  interpret molecular observations in large scale surveys. However, we must keep in mind that our cuts avoid  protostars and the hot cores or hot corinos, bipolar outflows, and HII regions formed around them. The binding energy of CS is estimated to be $\sim$3200$\pm$960 K \citep{Wakelam2017}, and therefore thermal evaporation is expected to be efficient when the dust temperature is > 50~K. Sputtering is also efficient as a mechanism to release molecules to the gas phase for shock velocities $>$ 10 km s$^{-1}$
\citep{Jimenez-Serra2008}. Therefore, high CS abundances are expected in hot cores/corinos and bipolar outflows where the icy mantles are eroded \citep{Bachiller1997, Jimenez-Serra2008,Esplugues2014, Crockett2014}.

\subsection{HCO$^+$, H$^{13}$CO$^+$, HC$^{18}$O$^+$}
\label{Subsec: HCO,H13CO,HC18O}

\begin{figure*}
\includegraphics[scale=.46]{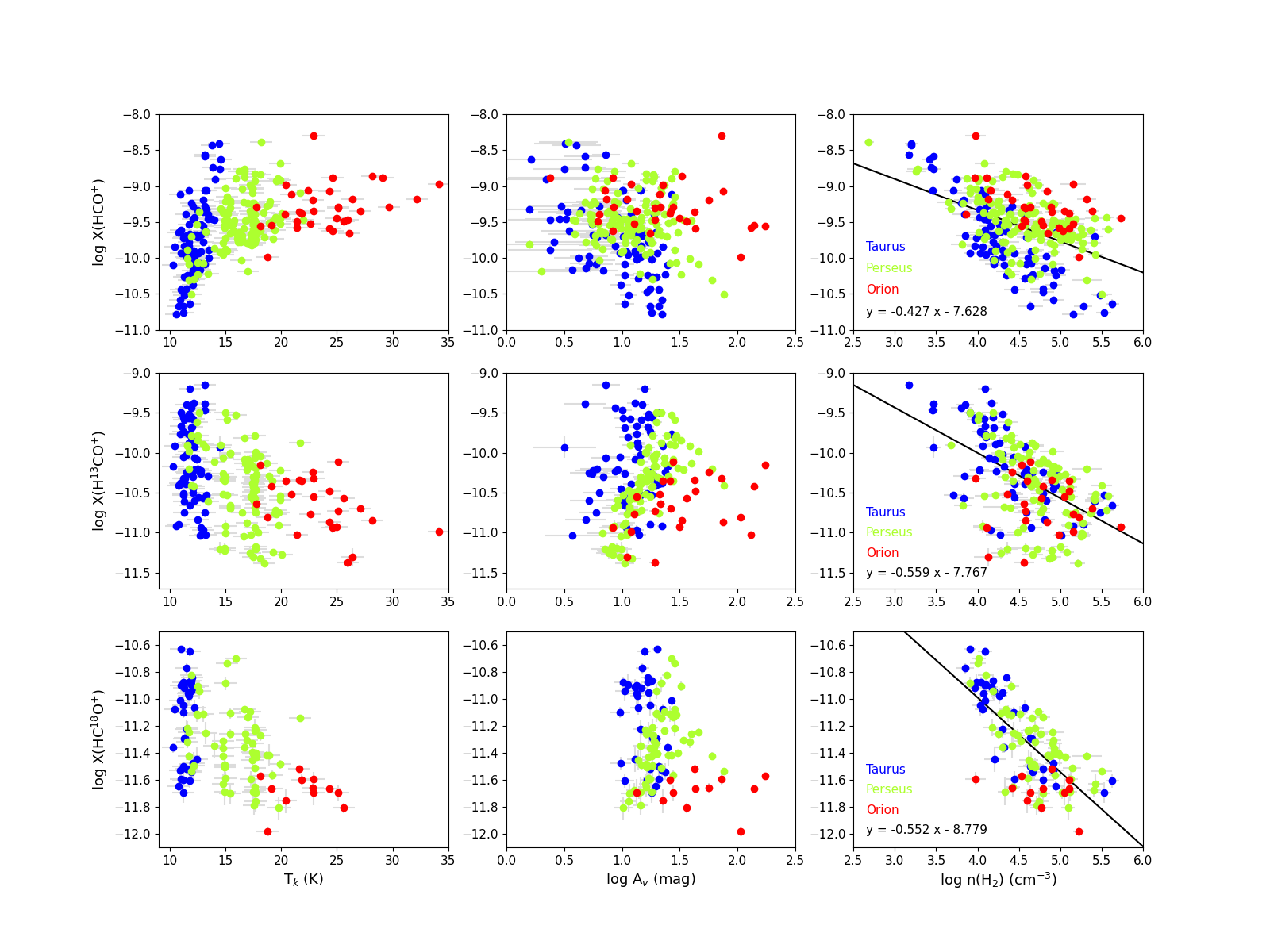}
\caption{Relation of the HCO$^+$ (top row), H$^{13}$CO$^+$ (middle row), and HC$^{18}$O$^+$ (bottom row) molecular abundances to cloud physical parameters: kinetic temperature (left column), extinction (middle column), and molecular hydrogen density (right column). The dataset is color-coded according to the molecular cloud of the points, as indicated in the legends. The black lines show the linear correlations between molecular abundances and molecular hydrogen density for the corresponding whole samples, with the correlation parameters indicated in the legends.}
\label{Fig: Molecular_abundances_HCOandfriends}
%\vspace{-0.1cm}
\end{figure*}

We derived the molecular abundances of HCO$^+$, H$^{13}$CO$^+$, and HC$^{18}$O$^+$ from the observations of their J =1$\rightarrow$0 rotational line. Isotopic $^{12}$C/$^{13}$C and $^{16}$O/$^{18}$O fractionation has been predicted under the physical conditions prevailing in molecular clouds for HCO$^+$ \citep{Roueff2015, Loison2019b, Colzi2020, Loison2020}. For this reason, we  prefer to independently fit the three isotopologs. The number of detections depends on the abundance of each isotopolog. The number of positions with A$_V$<8 mag detected in HCO$^+$ 1$\rightarrow$0 is significantly larger than those detected in H$^{13}$CO$^+$ 1$\rightarrow$0 (see Fig.~\ref{Fig: Molecular_abundances_HCOandfriends}). The HC$^{18}$O$^{+}$ 1$\rightarrow$0 line is only detected for A$_V > $8 mag. Thus, the observation of the three isotopologs allows us to probe different regions of the cloud. 

We explore the possible (anti-)correlations of the abundances of these ions with kinetic temperature, extinction, and molecular hydrogen density in Fig.~\ref{Fig: Molecular_abundances_HCOandfriends}. The abundance of the main isotopolog, X(HCO$^+$), seems to increase with temperature. 
This apparent correlation is the consequence of the fact that the HCO$^+$ 1$\rightarrow$0 line is optically thick. In this regime, line intensities simply reflect the excitation temperature in the layer with $\tau \sim$ 1, and our calculations  only provide a lower limit to the real HCO$^+$ abundance. Moreover, this line is self-absorbed in many positions, especially in Taurus \citep{Fuente2019}. This apparent correlation disappears for the less abundant isotopologs, H$^{13}$CO$^+$ and HC$^{18}$O$^+$, whose emission is expected to be optically thin. These isotopologs can be used as proxies for HCO$^+$. Figure \ref{Fig: Molecular_abundances_HCOandfriends} shows the decrease in the abundance of these molecular ions with density. This anti-correlation is particularly clear for HC$^{18}$O$^+$, that is, for the inner layers of the gas, as we only detect this molecule for A$_V$ > 8~mag.

\begin{figure*}
\includegraphics[scale=.37]{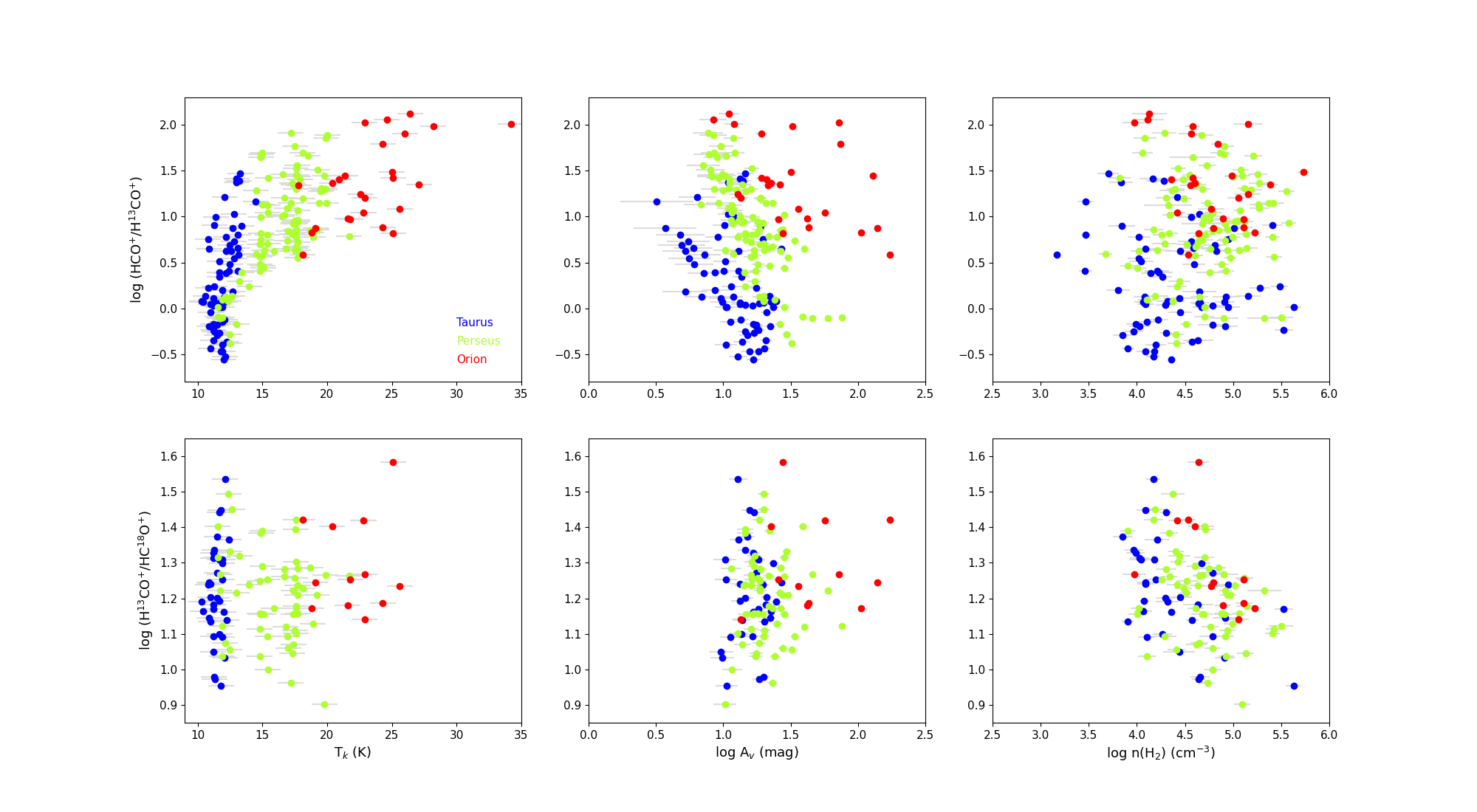}
\caption{Relation of the HCO$^+$/H$^{13}$CO$^+$ (top row) and H$^{13}$CO$^+$/HC$^{18}$O$^+$ (bottom row) molecular abundance ratios to cloud physical parameters: kinetic temperature (left), extinction (middle), and molecular hydrogen density (right). The dataset is color-coded according to the molecular cloud of the points, as indicated in the legend.}
\label{Fig: Molecular_abundances_HCO_comparative}
%\vspace{-0.1cm}
\end{figure*}

In Fig. \ref{Fig: Molecular_abundances_HCO_comparative} we show N(HCO$^+$)/N(H$^{13}$CO$^+$) and N(H$^{13}$CO$^+$)/N(HC$^{18}$O$^+$) as derived from their  J=1$\rightarrow$0 spectra. We obtain N(H$^{13}$CO$^+$)/N(HC$^{18}$O$^+$)$\approx$10$-$25, which is consistent with our assumption that the emission of these species is optically thin. It should be noted that the values of N(H$^{13}$CO$^+$)/N(HC$^{18}$O$^+$) are higher than approximately 10 for most positions, with an average value of about 15. This value is higher than that expected when taking into account the adopted $^{12}$C/$^{13}$C and $^{16}$O/$^{18}$O isotopic ratios. One reason is the presence of several velocity components at many of the observed positions. Only the most intense of these components is detected in N(HC$^{18}$O$^+$), which introduces uncertainties of a factor of two in the measured N(H$^{13}$CO$^+$)/N(HC$^{18}$O$^+$) ratio. Carbon isotopic fractionation could also explain the high values of the 
N(H$^{13}$CO$^+$)/N(HC$^{18}$O$^+$) ratio. Chemical models predict that the HCO$^{+}$/H$^{13}$CO$^+$ abundance ratio varies during cloud evolution and could be $<$60 for ages of a few 0.1 to $\sim$1 Myr for typical conditions in dark clouds \citep{Roueff2015, Colzi2020, Loison2020}. The overabundance of H$^{13}$CO$^+$ could push the N(H$^{13}$CO$^+$)/N(HC$^{18}$O$^+$) to values around $>$10.

The N(HCO$^+$)/N(H$^{13}$CO$^+$) ratio correlates with the gas temperature and is lower than the $^{12}$C/$^{13}$C isotopic ratio at all positions except a few points in Orion and Perseus. This strongly supports the interpretation that the HCO$^+$ 1$\rightarrow$0 line is optically thick at  all the positions in which H$^{13}$CO$^+$ is detected.  The obtained values of N(HCO$^+$) are only reliable at low visual extinction (A$_V
<$ 5 mag). At these low extinctions, we measured X(HCO$^+$) $\sim$ 10$^{-10}-$3$\times$10$^{-9}$. The HCO$^+$ abundance in the diffuse medium has been estimated to be N(HCO$^+$)/N(H$_2$)=3$\times$10$^{-9}$ \citep{Liszt2010, Liszt2016}, which is coincident with the upper end of our range.

The abundance ratio N(HCO$^+$)/N(CO) is traditionally used to estimate the ionization degree, X(e$^-$), in molecular clouds
\citep{Wootten1979, Guelin1982, Caselli1998, Zinchenko2009, Fuente2016, Fuente2019}. For regions where CO is not heavily depleted (constant CO abundance), X(HCO$^+$) $\propto$ $\frac{\zeta(H_2)}{n(H_2)X(e^-)}$ (see e.g., \citealp{Caselli1998}). If we assume that the cosmic ray molecular hydrogen ionization rate, $\zeta(H_2)$, remains constant for the visual extinctions A$_V >$ 8 mag probed with our HC$^{18}$O$^+$ data, the correlation found between HC$^{18}$O$^+$ and n(H$_2$) is consistent with $X(e^-)\propto$ n$^{-0.5}$, which agrees with chemical predictions for starless cores \citep{Caselli1998}. However, it should be noted that the Orion points lie systematically below the straight line fitted in Fig.~ \ref{Fig: Molecular_abundances_HCOandfriends} while positions in Taurus are systematically above this line. Although suggestive of a different X(e$^-$), a multitransition study of HCO$^+$ and its isotopologs and chemical modeling are required to establish firm conclusions.  A combination of uncertainties in n(H$_2$) in the low-density positions in Taurus and opacity effects due to the higher extinction towards the high-density positions towards Orion could also produce this effect.  

\subsection{HCN, H$^{13}$CN, HNC}
\label{Subsec: H13CN}

\begin{figure*}
\includegraphics[scale=.46]{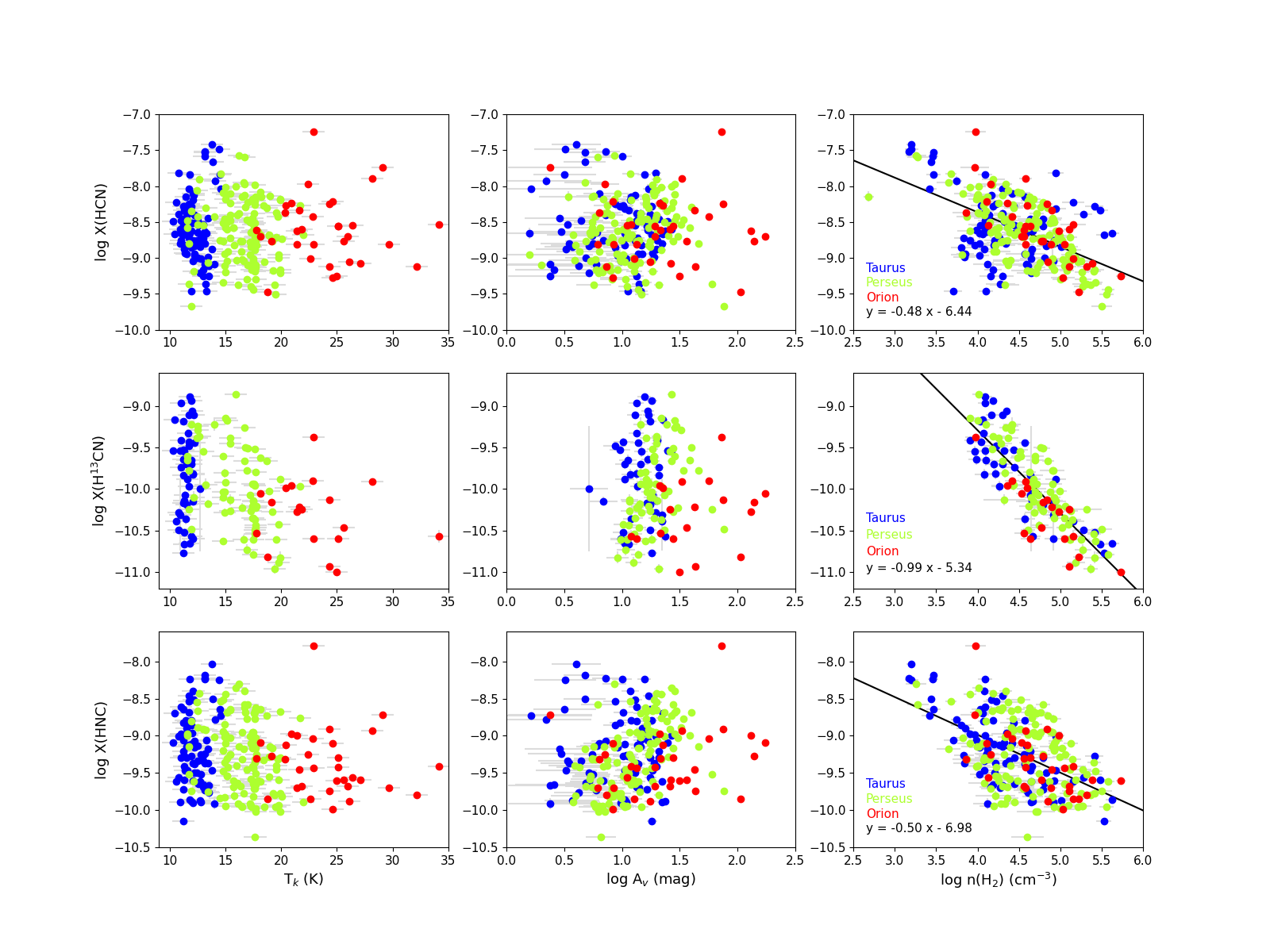}
\caption{Relation of the HCN (top row), H$^{13}$CN (middle row), and HNC (bottom row) molecular abundances to cloud physical parameters: kinetic temperature (left), extinction (middle), and molecular hydrogen density (right). The dataset is color-coded according to the molecular cloud of the points, as indicated in the legend. Black lines show the linear correlations between molecular abundances and molecular hydrogen density for the corresponding whole samples, with the correlation parameters indicated in the legends.}
\label{Fig: Molecular_abundances_H13CN}
%\vspace{-0.1cm}
\end{figure*}

The isomers hydrogen cyanide (HCN) and hydrogen isocyanide (HNC) are widely observed in the interstellar medium. They have been detected in diffuse clouds \citep{Liszt2001}, translucent molecular clouds \citep{Turner1997}, dark cloud cores \citep{Hirota1998, HilyBlant2010}, and in low-mass and massive star-forming regions \citep{Colzi2018, Goicoechea2019, Hacar2020, Legal2020}. Understanding their chemical behavior in all these environments is crucial for the correct interpretation of molecular observations. Figure \ref{Fig: Molecular_abundances_H13CN} shows the relationships between HCN and H$^{13}$CN molecular abundances and the physical parameters of the gas, classifying the data as a function of the hosting molecular cloud. The HCN column density was derived from the total area of the 1$\rightarrow$0 line (summing up of the hyperfine components) and assuming the same physical conditions as above. The HCN 1$\rightarrow$0 line is optically thick in almost all positions with A$_V$ > 8 mag. Moreover, the profile of the main hyperfine component presents deep self-absorption features towards the Taurus dark cores (see \citealp{Fuente2019}). This is the reason for the apparent low HCN abundance in Taurus positions. This kind of self-absorbed profile is less common in Perseus and does not appear in our limited sample of positions in Orion. 

In order to avoid opacity and self-absorption effects, we prefer to trace the behavior of this molecule using its isotopolog, H$^{13}$CN. We are aware that fractionation might be important for HCN and that the HCN/H$^{13}$CN ratio can vary by a factor of about two \citep{Roueff2015, Zeng2017, Loison2019a, Colzi2020, Loison2020}. Thus, H$^{13}$CN is a good proxy for HCN with an uncertainty of a factor of two. The H$^{13}$CN 1$\rightarrow$0 line is only detected in positions with A$_V >$ 8 mag (see Fig. \ref{Fig: Molecular_abundances_H13CN} ). We observe a strong anti-correlation between the molecular abundance and H$_2$ density, with less scatter than in the case of H$^{13}$CO$^+$ and HC$^{18}$O$^+$. The trend is common to the three molecular clouds, and remains for around two orders of magnitude in density. This similarity between HCO$^+$ and HCN isotopologs suggests a related chemistry in dark clouds. The recombination of protonated compound HCNH$^+$ is the most important formation mechanism for HCN and HNC in cold dark clouds \citep{Loison2014}.  The radical CN is also a product of this recombination which itself can react with H$_3^+$ to produce HCN$^+$ which is rapidly recycled to HCNH$^+$ by reactions with H$_2$. The abundance of these isomers is therefore sensitive to variations in the gas ionization degree and H$_3^+$ abundance, similarly to the case of the molecular ions H$^{13}$CO$^+$ and HC$^{18}$O$^+$.  \citet{Zinchenko2009} carried out a survey of HN$^{13}$C and H$^{13}$CN in massive star forming regions, and found an anti-correlation between the HN$^{13}$C abundance and the gas ionization degree. However, they did not obtain the same result for H$^{13}$CN. In contrast to HNC, the HCN abundance is boosted in the shocked gas associated to bipolar outflows and the hot cores in the early stages of massive star formation \citep{Schilke1992, Schoier2002,Rolffs2011,James2020, Hacar2020}.  
Our data show that the H$^{13}$CN abundance decreases with n(H$_2$) in the quiescent gas. \citet{HilyBlant2010} studied the chemistry of HCN, HNC, and CN in starless cores, deriving X(H$^{13}$CN)$\sim$10$^{-11}$ towards the millimeter emission peaks. These values are in good agreement with the values we obtained for densities n(H$_2$)$>$10$^5$~cm$^{-3}$, in the lower end of the observed range of H$^{13}$CN fractional abundances.

It is interesting to compare the behavior of X(H$^{13}$CN) with that of X(HNC), whose relationships with cloud physical parameters are also plotted in Fig. \ref{Fig: Molecular_abundances_H13CN}. X(HNC) also presents a decreasing abundance with molecular hydrogen density, but with a shallower power law of slope $\sim-$0.5. This could be mainly due to the opacity, because the HNC line is expected to be optically thick in dense cores. This shallow slope is also observed in the X(HCN) versus n(H$_2$) correlation.

\begin{figure}
\centering
\includegraphics[scale=.46]{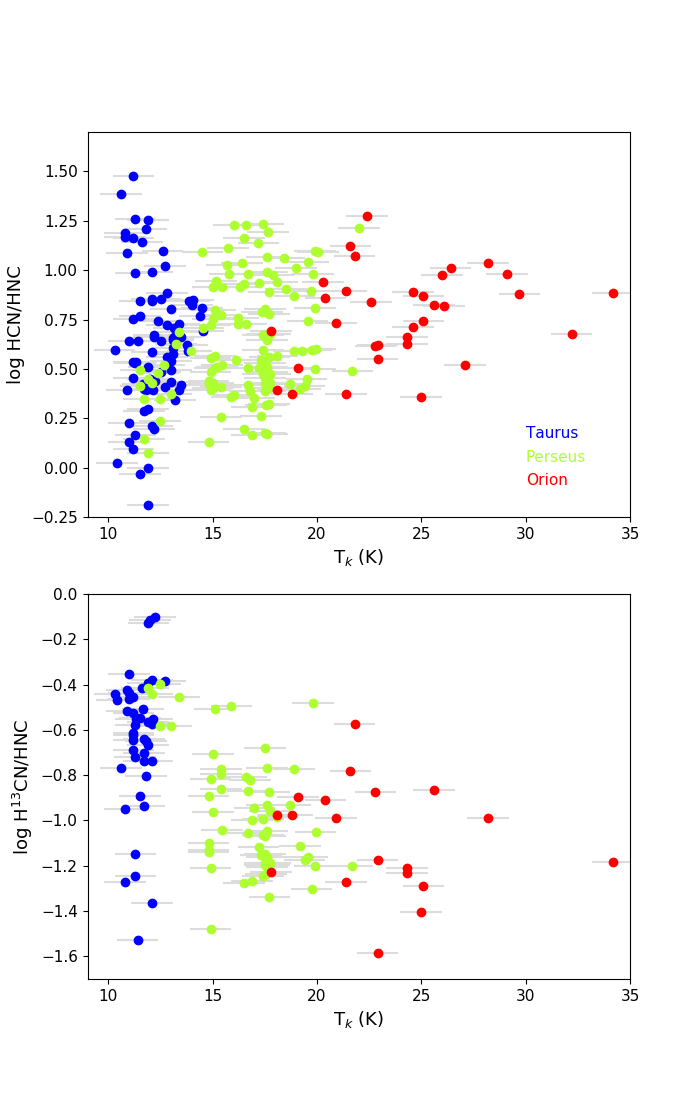}
\caption{Relation of the HCN/HNC (top) and H$^{13}$CN/HNC (bottom) line ratios to kinetic temperature. The dataset is color-coded considering the molecular cloud of the points, as indicated in the legend.}
\label{Fig: Molecular_abundances_ratHCNH13CN_HNC}
%\vspace{-0.1cm}
\end{figure}

The HCN/HNC line ratio has been proposed as a direct probe of the gas kinetic temperature \citep[e.g.,][]{Goldsmith1981,Goldsmith1986,Baan2008,Hacar2020}, with increasing values of this ratio probing warmer material, which could lead to it being used as a new chemical thermometer of the molecular interstellar medium. In order to explore this proposal, Fig. \ref{Fig: Molecular_abundances_ratHCNH13CN_HNC} shows the relationships between the HCN/HNC and H$^{13}$CN/HNC line ratios with kinetic temperature. In our case, the lack of reliability of derived HCN abundances due to the mentioned self-absorption effects in Taurus and partially in Perseus prevents a reliable analysis of the trends found in these clouds. On the other hand, Orion points indeed show slight increase in this line ratio with temperature, in good agreement with results by \citet{Hacar2020}, but the number of points is not high enough to give a statistically robust conclusion. The observed trend for Orion points disappears when considering the H$^{13}$CN/HNC line ratio. Indeed, considering all the points (Taurus, Perseus and Orion), the H$^{13}$CN/HNC ratio seems to decrease.  Again, we  think that this is not real, and is the effect of the high optical  depths  of   the HNC J=1$\rightarrow$0 line in cold cores (see, e.g., \citealp{Daniel2013}), which lead to an underestimation of N(HNC). Multi-transition observations of HCN, HNC, and their  isotopologs H$^{13}$CN and HN$^{13}$C are  required to constrain the X(HCN)/X(HNC) ratio. A deeper analysis based on chemical models and 3D radiative transfer calculations should be carried out in order to determine the behavior of this abundance ratio with kinetic temperature.

\subsection{H$_2$S}
\label{Subsec: H2S}

\begin{figure*}
\includegraphics[scale=.41]{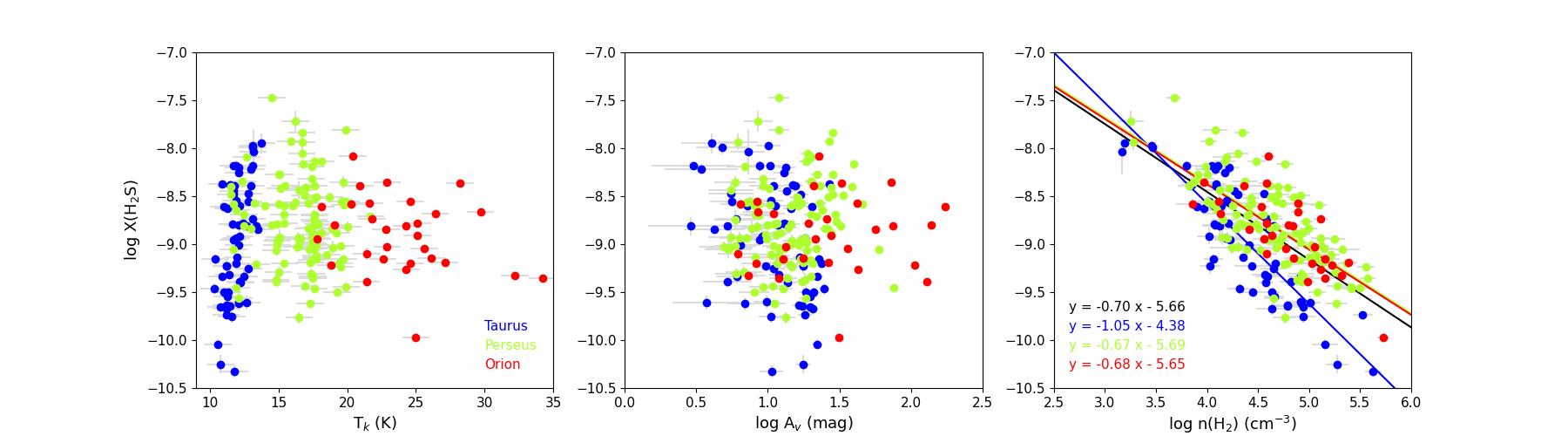}
\caption{Relation of the H$_2$S molecular abundance to cloud physical parameters: kinetic temperature (left), extinction (middle), and molecular hydrogen density (right). The dataset is color-coded according to the molecular cloud of the points, as indicated in the legend. Lines in the right plot show the linear correlations between X(H$_2$S) and n(H$_2$) for the whole sample, in black, and for each one of the molecular clouds, color-coded as in the points. The corresponding correlation parameters are indicated in the legend. }
\label{Fig: Molecular_abundances_H2S}
%\vspace{-0.1cm}
\end{figure*}

Knowledge of the H$_2$S molecule is key to understanding sulfur chemistry. Sulfur is one of the most abundant elements in the universe, and its presence in the interstellar medium has been widely studied. Sulfur atoms in interstellar ice mantles are expected to preferentially form H$_2$S because of the high hydrogen abundances and the mobility of hydrogen in the ice matrix. Therefore, studying H$_2$S abundance in the gas phase is essential to our understanding of the chemical processes that lead to sulfur depletion in these environments. Of particular importance, this molecule was the subject of a previous paper of the GEMS series, where the H$_2$S observations of TMC 1-C and Barnard 1b were analyzed \citep{Navarro2020}. In \citet{Navarro2020} we estimated the o-H$_2$S abundance using the ortho-H$_2$O collisional coefficients calculated by \citet{Dubernet2009}, scaled to ortho-H$_2$S. Here we use the specific o-H$_2$S collisional coefficients with o-H$_2$ and p-H$_2$ recently calculated by \citet{Dagdigian2020}. The difference between these two sets of collisional coefficients is of a factor of $\sim$1.7 at 10~K in collisions with p-H$_2$, which implies that the re-estimated ortho-H$_2$S column densities in TMC 1 and B1b are a factor of $\sim$1.7 lower than those reported by \citet{Navarro2020}. This moderate factor is within the usual uncertainties of column density estimates and does not affect any of the conclusions of this paper. The ortho-H$_2$S abundances derived from the observations of the J$_{K_1,K_2}$=1$_{1,0}\rightarrow1_{0,1}$ line are shown in Fig. \ref{Fig: Molecular_abundances_H2S}. These abundances are represented as a function of the gas physical parameters, and the points are classified according to their molecular clouds. We also observe in this case a strong anti-correlation between H$_2$S abundance and molecular hydrogen density, spanning over around three orders of magnitude. Interestingly, we do observe a displacement between the trend of the points belonging to Taurus and that shown by the warmer Perseus and Orion. This behavior was already observed by \citet{Navarro2020} based on the analysis of the data towards TMC~1-C and Barnard B1b. The analysis of the whole sample confirms and extends this behavior, with the observed positions in Taurus following a steeper law with density than those in Perseus and Orion. \citet{Navarro2020} explained this behavior as the consequence of the different chemical desorption efficiency in bare and ice-coated grains. 

Simultaneously with H$_2$S, we observed the H$_2^{34}$S  J$_{K_1,K_2}$=1$_{1,0}\rightarrow1_{0,1}$ line with very few detections, which proves the low optical depth of the same line of the main isotopolog. Optical depth effects are therefore not expected to bias the (anti-)correlation with density.

\subsection{SO, $^{34}$SO}
\label{Subsec: SO,34SO}

One of the most abundant and easily observable species is SO which, along with CS, is the most abundant S-bearing molecule in the gas phase. Within GEMS, we observed the SO J$_N$=2$_{2}\rightarrow1_{1}$, 2$_{3}\rightarrow1_{2}$, 3$_{2}\rightarrow2_{1}$, 3$_{4}\rightarrow2_{3}$, and 4$_{4}\rightarrow3_{3}$ lines, allowing a multi-transition study. However, in most positions we only detected one or two of the observed lines. In order to apply a uniform analysis method to all the positions included in this statistical analysis, we estimated the SO column density based on the intensities of the SO J$_N$ 2$_{2}\rightarrow1_{1}$ line, assuming the molecular hydrogen density estimated from CS. In addition to the main isotopolog, we observed the  J$_N$=2$_{3}\rightarrow1_{2}$ line of the rarer isotopolog $^{34}$SO which can provide important information with which to constrain the SO abundance in the most obscured regions. 

\begin{figure*}
\includegraphics[scale=.37]{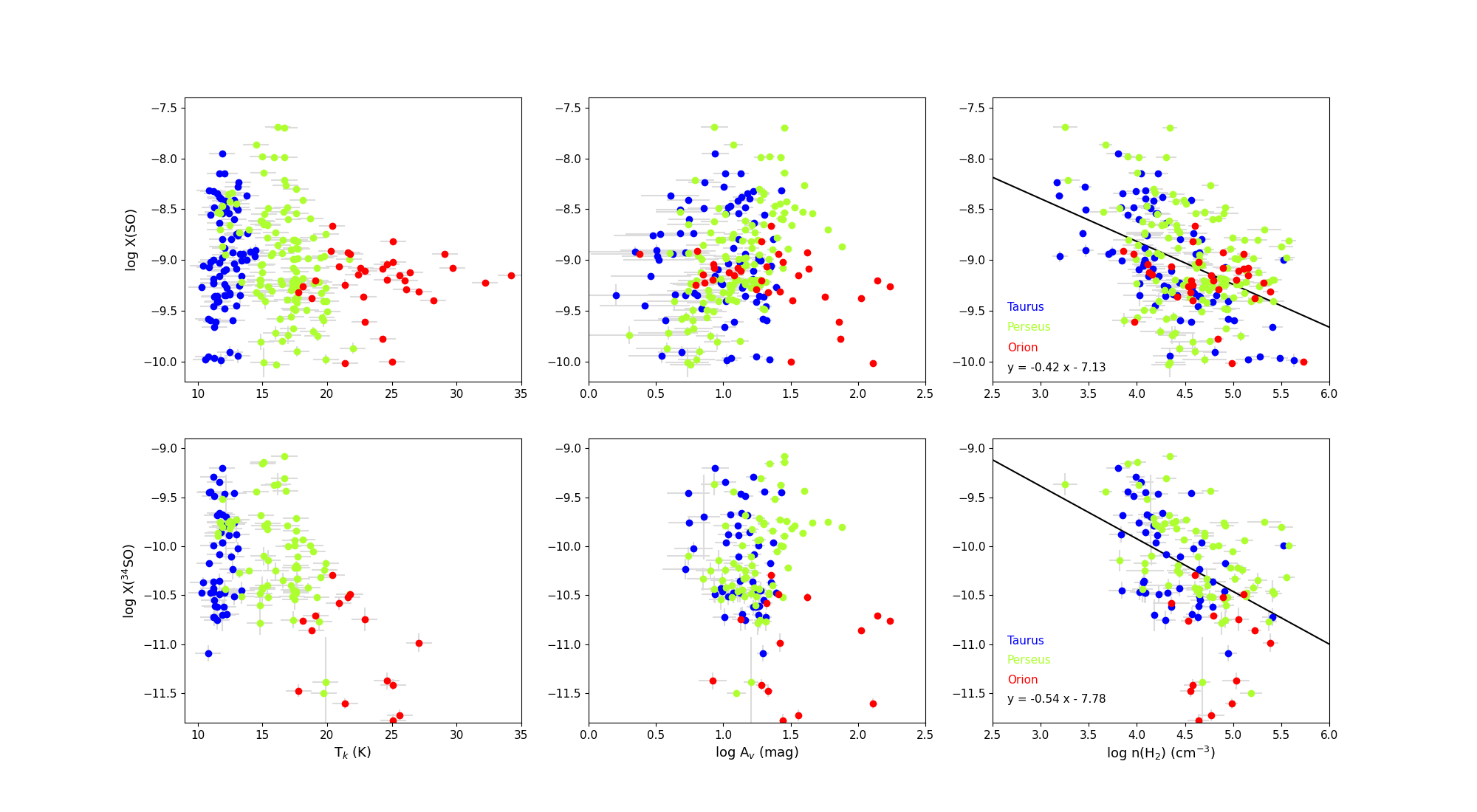}
\caption{Relation of the SO (top row) and $^{34}$SO (bottom row) molecular abundances to cloud physical parameters: kinetic temperature (left), extinction (middle), and molecular hydrogen density (right). The dataset is color-coded according to the molecular cloud of the points, as indicated in the legend. Black lines show the linear correlations between molecular abundances and molecular hydrogen density for the corresponding whole samples, with the correlation parameters indicated in the legends.}
\label{Fig: Molecular_abundances_SO_34SO}
%\vspace{-0.1cm}
\end{figure*}

\begin{figure}
\centering
\includegraphics[scale=.38]{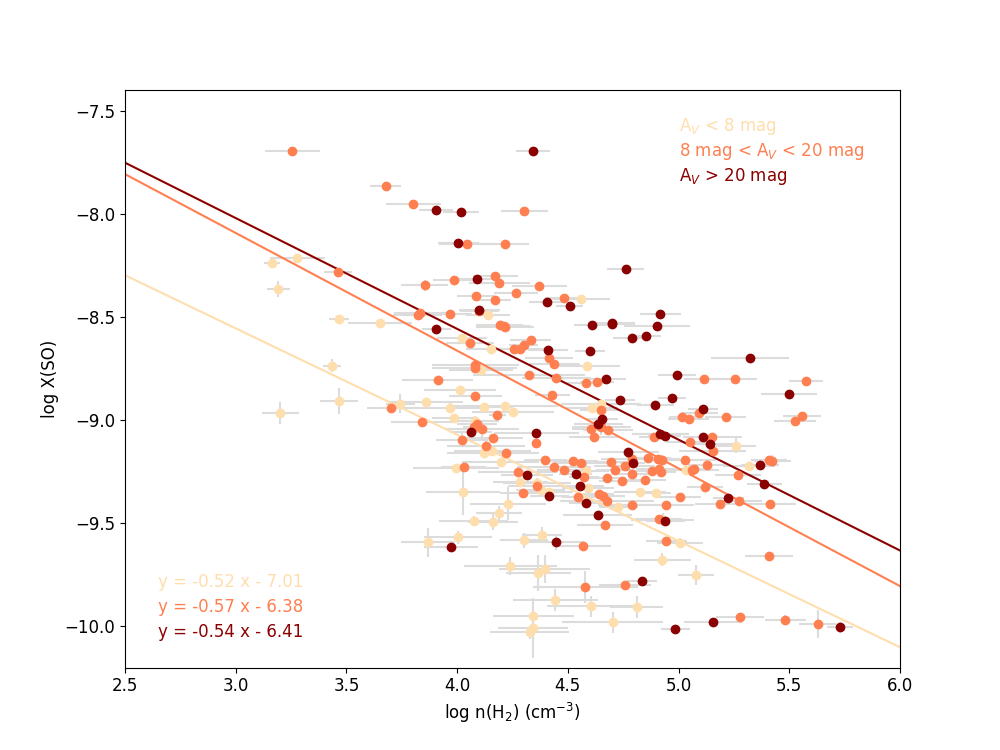}
\caption{Relation of the SO molecular abundance to the molecular hydrogen density. The dataset is color-coded considering bins of extinction, as indicated in the legend. Lines show the linear correlations between molecular abundances and molecular hydrogen density for each extinction bin, with the correlation parameters indicated in the legend.}
\label{Fig: Molecular_abundances_SO_34SO_n_ext}
%\vspace{-0.1cm}
\end{figure}

The relations between SO and $^{34}$SO molecular abundances and gas physical parameters are shown in Fig.~\ref{Fig: Molecular_abundances_SO_34SO}, with points classified as a function of their molecular clouds. As in previous cases, a linear decreasing relation with molecular hydrogen density is found in both cases, but for SO isotopologs the relation seems to be weaker and with a wider scatter. In order to explore the origin of this scatter, we plot the relationship between SO molecular abundance and n(H$_2$)  in  Fig. \ref{Fig: Molecular_abundances_SO_34SO_n_ext}, with points classified according to bins of extinction, and the corresponding linear relations existing for each case. We observe that the slope is similar for all ranges of extinction, but the intercept changes, being lower for A$_V<$ 8 mag. This suggests that the abundance of SO is lower in the outer part of the cloud, more likely because of the enhanced UV radiation.  

There are some positions with high N(SO)/N($^{34}$SO) ($>$50), and most of these positions are in Orion. We do not expect isotopic fractionation for $^{32}$S/$^{34}$S. Therefore, we discuss some observational uncertainties. First of all, many positions of the cut Ori-C2-1 present two velocity components in the SO line. Only the most intense of these components are detected in $^{34}$SO. As we did not carry out a differentiated study for each velocity component, N(SO)/N($^{34}$SO) is expected to be overestimated towards these positions. However, it is difficult to justify N(SO)/N($^{34}$SO) $>$50 based only on this effect. We recall that we assume the same physical conditions for SO and $^{34}$SO. A steep gradient in temperature and density along the line of sight could also induce anomalously high N(SO)/N($^{34}$SO) ratios.

\subsection{HCS$^+$, OCS}
\label{Subsec: HCS,OCS}

\begin{figure*}
\includegraphics[scale=.37]{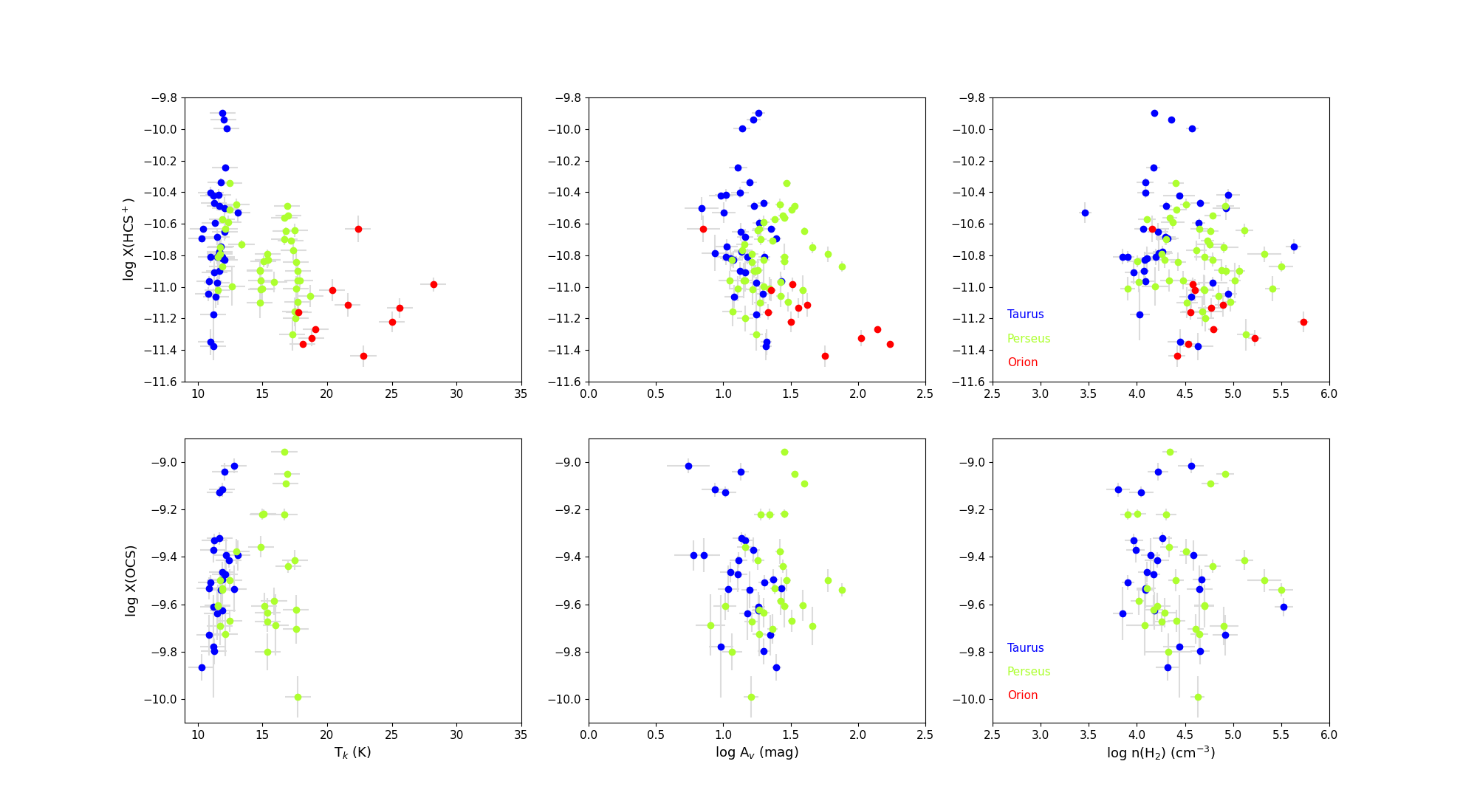}
\caption{Relation of the HCS$^+$ (top row) and OCS (bottom row) molecular abundances to cloud physical parameters: kinetic temperature (left), extinction (middle), and molecular hydrogen density (right). The dataset is color-coded according to the molecular cloud of the points, as indicated in the legend.}
\label{Fig: Molecular_abundances_HCS_OCS}
%\vspace{-0.1cm}
\end{figure*}

The S-bearing species HCS$^+$ and OCS are detected only in a fraction of the observed positions, meaning lower statistical significance. Nevertheless, we consider that it is interesting to include them in our study. Figure \ref{Fig: Molecular_abundances_HCS_OCS} shows the relation between the HCS$^+$ and OCS molecular abundances and the physical parameters of the clouds. Both species are only detected for extinctions greater than $\sim$ 8 mag, and OCS is only detected in Taurus and Perseus, but not in Orion. 
Based on our data, we derive X(OCS) $<$ 4$\times$10$^{-10}$ in Ori-C1-2, X(OCS) $<$ 1$\times$10$^{-9}$ in Ori-C2-2, and X(OCS) $<$ 5$\times$10$^{-10}$ in Ori-C1-3, which are the extinction peaks of the three cuts observed in Orion. These upper limits are in the range of OCS abundances derived in Perseus and Taurus. Therefore, the nondetection might be the consequence of the limited sensitivity of the Orion observations.
 
HCS$^+$ is the protonated species of CS and its precursor in diffuse clouds and the external layers of the photon-dissociation region \citep{SternbergDalgarno1995, Lucas2002, Riviere2019}. Contrary to HCO$^+$, deeper into the molecular cloud HCS$^+$ is rapidly destroyed by reaction with O, and CS is formed by neutral--neutral reactions such as SO + C  $\rightarrow$ CS +O. In agreement with its origin in the outer layers of molecular clouds, in  Fig.~\ref{Fig: Molecular_abundances_HCS_OCS} we observe a steep decrease of the HCS$^+$ with visual extinction. In fact, contrary to most of the species studied, the (anti-)correlation of X(HCS$^+$) with A$_V$ presents less scattering than its relationship with gas density. In the case of OCS, we do not find any clear relation between its abundance and the physical conditions of the clouds. This is most likely due to the small number of detections, which does not allow us to properly cover the parameter 
space of our study.

\subsection{$^{13}$CO, C$^{18}$O}
\label{Subsec: 13CO,C18O}

We derived the $^{13}$CO and C$^{18}$O abundances from the observations of their J=1$\rightarrow$0 rotational lines. The major uncertainties in our calculations come from opacity effects in the studied lines and the assumption that gas and dust share the same temperature. Assuming a standard CO abundance, X(CO)$\sim$8$\times$10$^{-5}$, and the same isotopic ratios as in previous sections, the $^{13}$CO 1$\rightarrow$0  and  C$^{18}$O 1$\rightarrow$0  lines are expected to be optically thin for A$_V$ < 8$-$10~mag and A$_V  <$ 60$-$100~mag respectively. Therefore the estimated C$^{18}$O abundances are reliable in almost the entire range of visual extinctions considered in our study, with the exception of a few positions with A$_V >$ 60 mag that have been removed from the plots and are not considered in this section hereafter. The second approximation is our assumption that T$_d$=T$_{gas}$. As mentioned in Sect.~\ref{Sec.Physical conditions}, this approximation is reasonable for well-shielded regions of molecular clouds, A$_V >$8 mag (see also  \citealp{Roueff2020}), where the $^{13}$CO 1$\rightarrow$0 line is expected to be optically thick; however,it is inaccurate at the cloud edges, where UV photons heat the gas more efficiently than the dust by the photoelectric ejection of electrons from grains and polycyclic aromatic hydrocarbons (PAHs). Caution should be taken in sources bathed in enhanced UV fields, like Orion and IC 348, where the gas temperature is $\sim$15~K higher than the dust temperature in the cloud surface. In order to estimate the error due to the uncertainty in the gas temperature in these regions, we carried out a simple calculation assuming typical physical values of n(H$_2$)=10$^4$ cm$^{-3}$ and N($^{13}$CO)$\sim$ a few 10$^{15}$ cm$^{-2}$. The difference in the derived N($^{13}$CO) assuming temperatures of between 15~K and 30~K is of $\sim$20\%. Our approximation is better for Taurus, where the incident UV field is low and gas and dust temperature agree within $<$ 5~K even in the more external layers (see Fig.~\ref{grains}). 

\begin{figure*}
\includegraphics[scale=.46]{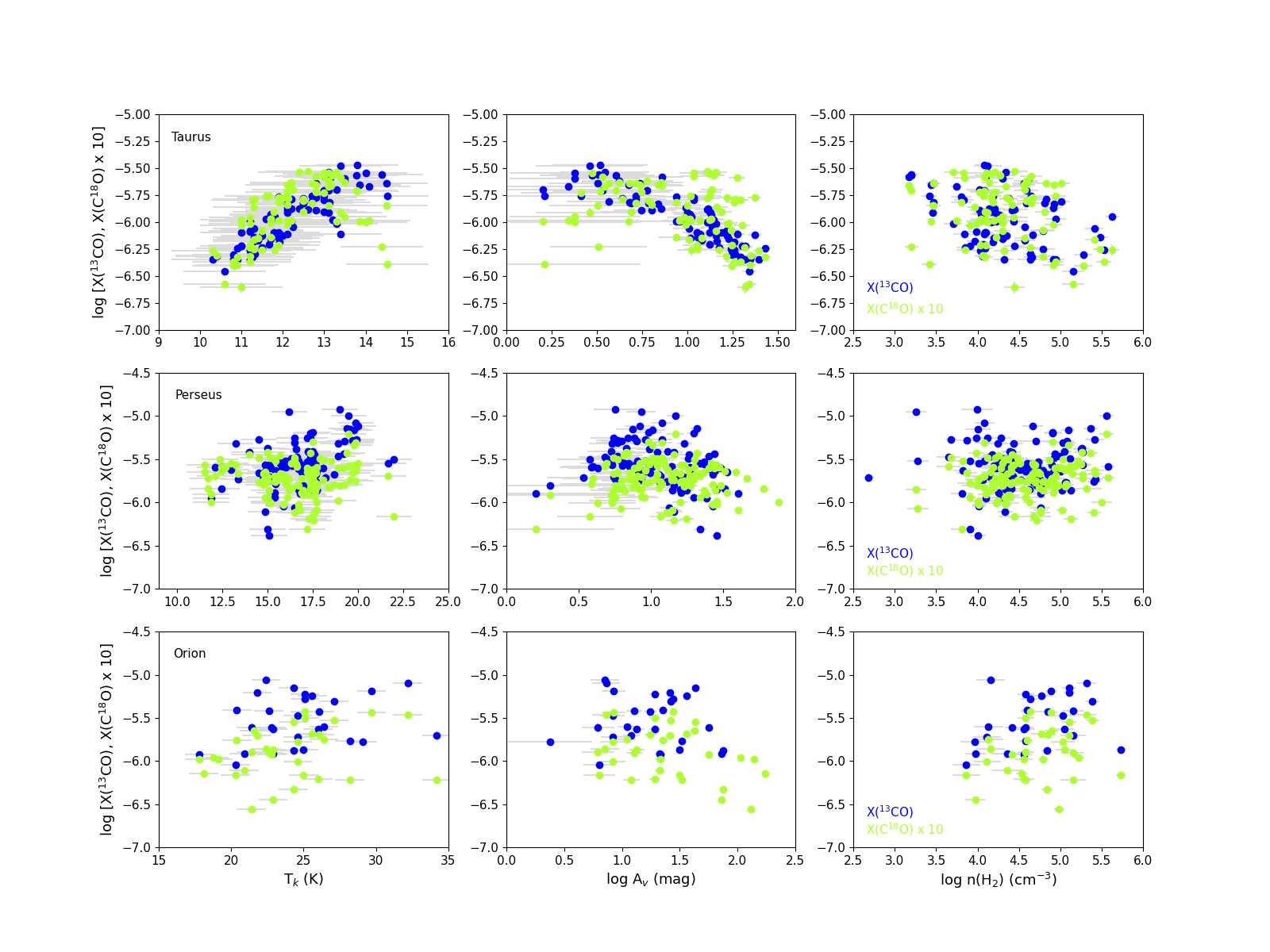}
\caption{Relation of the $^{13}$CO (blue) and C$^{18}$O (green, multiplied by a factor of ten) molecular abundances to cloud physical parameters: kinetic temperature (left), extinction (middle), and molecular hydrogen density (right), for the three observed molecular clouds, Taurus (top row), Perseus (middle row), and Orion (bottom row).}
\label{Fig: Molecular_abundances_13CO_C18O}
%\vspace{-0.1cm}
\end{figure*}

The relationship between the derived abundances and the physical parameters of the clouds is represented in Fig. \ref{Fig: Molecular_abundances_13CO_C18O}, where the positions of Taurus, Perseus, and Orion are represented in different panels. Points belonging to Taurus show clear trends with temperature and visual extinction, and a poorer correlation with density. The $^{13}$CO abundance increases with visual extinction until A$_V \sim$ 3 mag; for A$_V >$ 3 mag, the $^{13}$CO abundance is decreasing with visual extinction. As mentioned above, the $^{13}$CO line is expected to be optically thick for A$_V >$ 8 mag, and the calculated $^{13}$CO abundance could be underestimated. For A$_V >$ 8 mag, the C$^{18}$O isotopolog is a better proxy for CO. The abundance peak of  C$^{18}$O is found at A$_V \sim$ 5 mag, that is, shifted with respect to the position of the X($^{13}$CO) peak. For A$_V>$ 5~mag, we observe a decreasing trend in the C$^{18}$O abundance with visual extinction, although the scatter is wider than in $^{13}$CO. The shift between the abundance peak of $^{13}$CO and C$^{18}$O is not due to optical depth effects but to an increase of the N($^{13}$CO)/N(C$^{18}$O) ratio at A$_V <$ 5 mag. This offset between the two isotopologs was already observed by \citet{Cernicharo1987} and \citet{Fuente2019}, and was interpreted as a consequence of the selective photodissociation and isotopic fractionation \citep{Fuente2019}.

There is a tight correlation between X($^{13}$CO) and X(C$^{18}$O) with temperature in Taurus. We observe an increasing relation between the abundances of these species with the gas kinetic temperature until T$_K \sim$ 14 K for $^{13}$CO and until T$_K \sim$ 12$-$13 K in the case of C$^{18}$O, and a decrease in the abundances for higher temperatures. In Taurus, the positions with T$_K >$ 14 K correspond to the outer layers of the cloud, where the molecules are photo-dissociated.We measure X(C$^{18}$O) = (1$-$5)$\times$10$^{-8}$ in the coldest cores. 

In Perseus we observe the same trends as in Taurus but with a wider scatter. Peaks of the observed distributions for both $^{13}$CO and C$^{18}$O are found at larger extinctions than in the case of Taurus (A$_V \sim$ 7 mag for $^{13}$CO and A$_V \sim$ 10 mag for C$^{18}$O), which is expected because the ambient UV field in Perseus is higher than in Taurus. Indeed, based on the Herschel dust temperature and extinction maps, \citet{Navarro2020} estimated $\chi_{UV}\sim$24 for Barnard 1b, which is located in a moderately active star-forming region in Perseus \citep{Hatchell2007a, Hatchell2007b}. 

We do not detect any clear trend in the Orion positions. It should be noted that dust temperatures are $>$ 15 K towards all positions in Orion. 
Nevertheless, the X(C$^{18}$O)$<$10$^{-7}$ towards some positions, supporting the interpretation that part of the CO might remain locked in grains for warmer temperatures.  \citet{Cazaux2017} analyzed the CO depletion from a microscopic point of view, finding that the CO molecules depleted on grain surfaces show a wide range of binding energies, from 300 to 830 K (T$_{evap}\sim$15$-$30~K), depending on the conditions (monolayer or multilayer regime) in which CO has been deposited. Low abundances of C$^{18}$O have also been found in the dense and warm inner regions of protostellar envelopes \citet{Yildiz2010, Fuente2012,Anderl2016}, where complex organic chemistry is going on. In general, we observe larger scatter in the correlations of X($^{13}$CO) and X(C$^{18}$O) with the physical parameters in Orion than in Taurus and Perseus. The complex structure of this giant molecular cloud and the fact that  it is located further than Perseus and Taurus produce a mix of regions with different physical conditions along the line of sight and within the beam of the 30m, increasing the scattering in the estimated abundances. Moreover, our approximation of deriving the molecular abundances assuming a single density and temperature is more uncertain.

\begin{figure*}
\includegraphics[scale=.41]{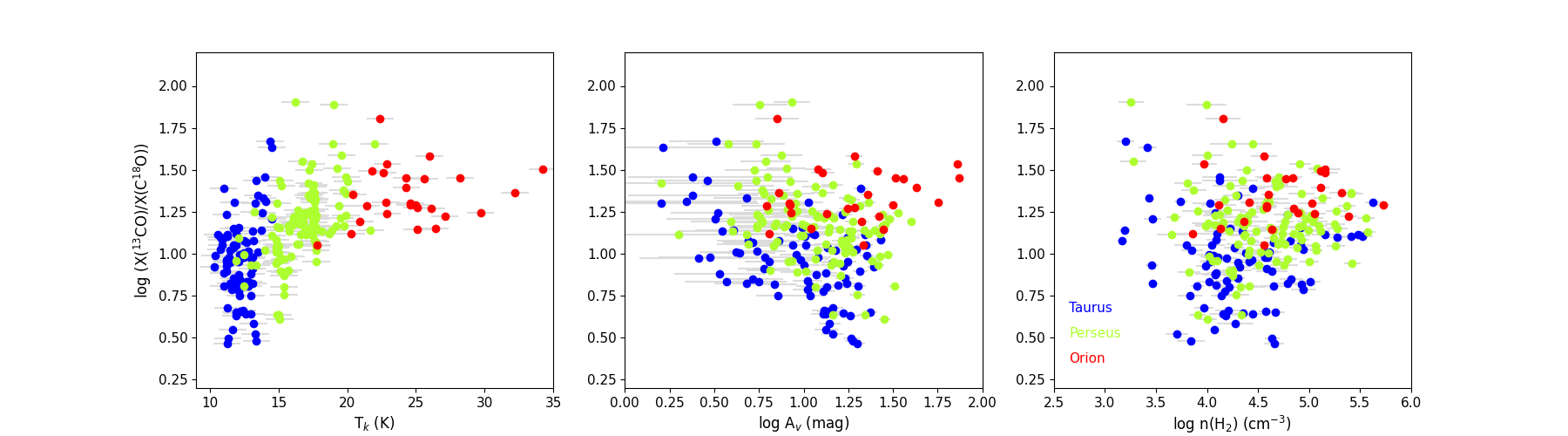}
\caption{Relation of the $^{13}$CO/C$^{18}$O molecular abundance ratio to cloud physical parameters: kinetic temperature (left), extinction (middle), and molecular hydrogen density (right). The dataset is color-coded considering the molecular cloud of the points, as indicated in the right-hand box.}
\label{Fig: Molecular_abundances_rat13COC18O}
%\vspace{-0.1cm}
\end{figure*}

Figure \ref{Fig: Molecular_abundances_rat13COC18O} represents the relations between N($^{13}$CO)/N(C$^{18}$O) and the gas physical parameters. In accordance with discussions above, we observe a decreasing relation between this ratio and extinction in the case of Taurus points, which can be understood as the consequence of an increase in the  $^{13}$CO 1$\rightarrow$0 line opacity with visual extinction. This trend is also reproduced in Perseus. Orion, on the other hand, does not show a significant trend along the selected cuts.  The mean values of N($^{13}$CO)/N(C$^{18}$O) are 10.9$\pm$7.5 in Taurus, 17.3$\pm$11.5 in Perseus, and 23.7$\pm$10.4 in Orion, i.e., higher in Orion than in Perseus and Taurus.

\section{Statistical correlations}
\label{Sec: Statistical correlations}

In this section,  we analyze the statistical trends shown in the dataset  in a quantitative and uniform way. In order to assess the degree of correlation observed between different parameters, we use the Pearson, Spearman, and Kendall correlation coefficients for the relations between the different parameters considered. These coefficients vary between values of +1 and -1, where the value indicates the strength of the relationship between the variables ($\pm$1 for a perfect degree of correlation and 0 for unrelated variables), and the sign indicates the direction of the relationship (positive for correlation and negative for anticorrelation). Pearson coefficient is the most widely used correlation parameter to measure the degree of linear relationship between variables. On the other hand, Spearman and Kendall coefficients measure the existence of  a monotonic, not necessarily linear, relationship between variables. The Kendall coefficient is usually more robust and is the most used in the case of small samples, but both are based on the same assumptions and provide useful information. Along with the coefficient value, the probability of the null hypothesis (no relation between variables) is also computed, indicating the strength of the result provided by the coefficient. Values for the probability of the null hypothesis higher than 0.001 are considered to indicate no relation at all. We consider the existence of a linear correlation between the corresponding magnitudes when the Pearson correlation coefficient is greater than or similar to $\sim$0.5 in absolute value, and a confirmed strong linear correlation when it is greater than or similar to $\sim$0.7 in absolute value. The relation is confirmed using the Spearman and Kendall coefficients as a further test: in the case of a linear relation, the Pearson and Spearman coefficients are similar and in good agreement with the Kendall coefficient, although the latter is systematically smaller in absolute value. Furthermore, a significantly greater value of the Spearman coefficient with respect to the Pearson coefficient in a particular case may indicate, if in good agreement with the corresponding Kendall coefficient, the existence of a nonlinear correlation.

\subsection{Relations between molecules}
\label{Subsec: Coef molecules}

\begin{figure*}
\centering
\includegraphics[scale=.45]{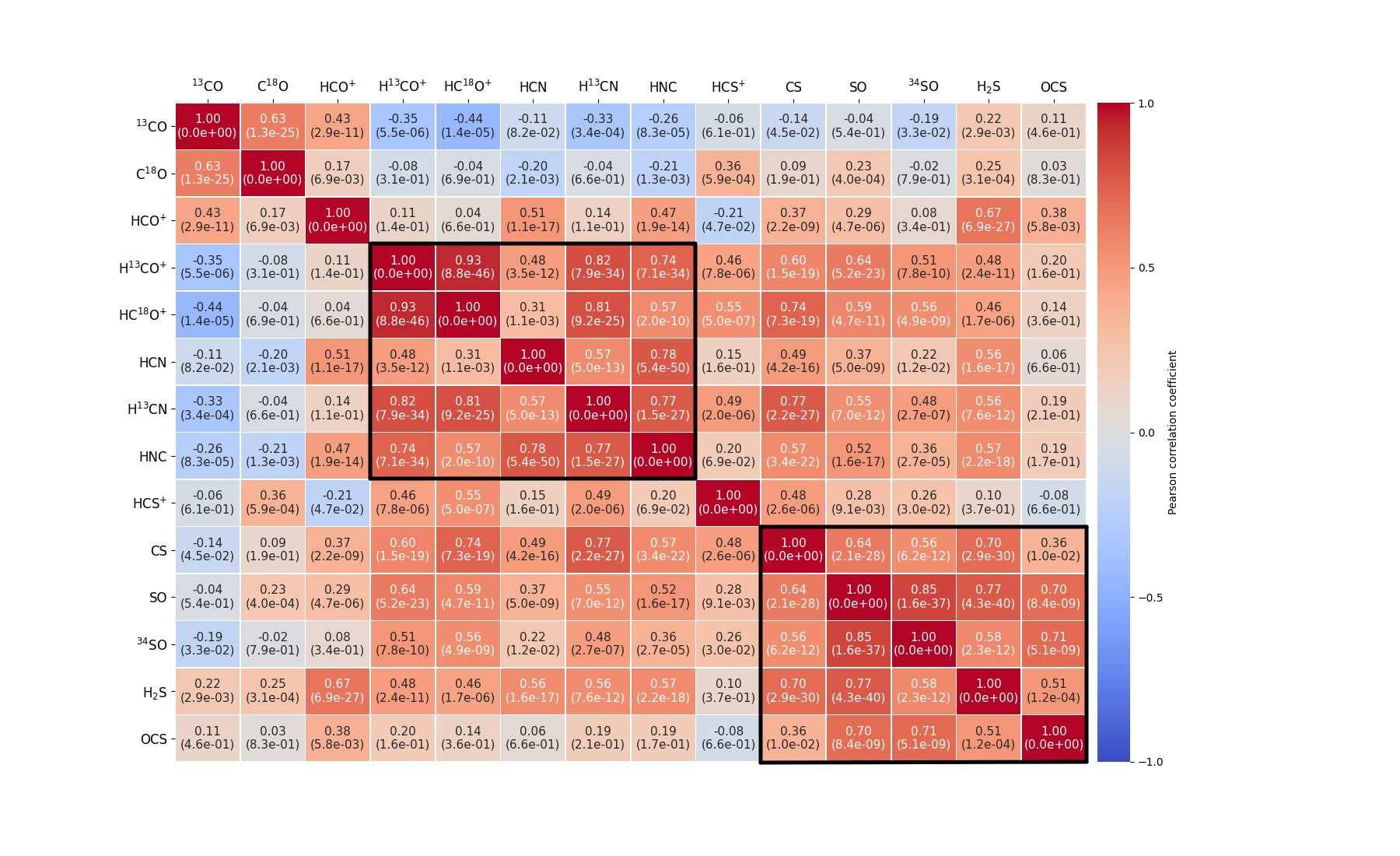}
\caption{Pearson correlation coefficients of the relations between the molecular abundances of the studied species. Numbers in brackets correspond to the null hypothesis probability in each case, i.e. probability of no relation between variables, where a value higher than 0.001 is considered to indicate no ralation at all. Black rectangles indicate groups of molecules with strong correlations (see text).}
\label{Fig: Coef_mol}
%\vspace{-0.1cm}
\end{figure*}

We explore the existence of behavioral similarities between the studied molecules by analyzing the relations between their derived molecular abundances. Figure \ref{Fig: Coef_mol} includes the computed Pearson coefficients for these relations. In this case, we do not include the Spearman and Kendall coefficients because the obtained relations are all linear and these coefficients do not significantly deviate from the Pearson coefficient value, and therefore do not provide additional information. Some strong correlations between groups of molecules are observed, as indicated in Fig.~\ref{Fig: Coef_mol}. In the first place, there is an almost complete correlation between H$^{13}$CO$^+$ and HC$^{18}$O$^+$, which is the expected behavior for isotopologs and confirms the results observed in Fig. \ref{Fig: Molecular_abundances_HCO_comparative}. These molecular ions  also correlate with H$^{13}$CN and HNC, and to a lesser extent with CS and HCS$^+$. It should be noted that HCN is not strongly correlated with H$^{13}$CN, which is a consequence of the high optical depth and the strong self-absorption features in the HCN 1$\rightarrow$0 spectra. As discussed in Sect. \ref{Sec: Molecular column densities}, the abundances of all these species steeply decrease with molecular hydrogen density. We propose molecular depletion on the grain surfaces and variations in the gas ionization degree as the causes of this behavior. Indeed, the protonated compounds HCO$^+$ and HCS$^+$ are rapidly destroyed by recombination with electrons to give back the neutral species, CO and CS. On the other hand, the neutral compound HCN is efficiently formed in dark clouds by the dissociative recombination of HCNH$^+$. These three latter ionic species are sensitive to the depletion on the grain surfaces of the neutral compounds, CO, CS, and HCN, respectively, but also to variations in X(e$^-$). In the absence of shocks and UV photons, cosmic rays are the main ionization agent. Cosmic rays are attenuated with N$_{\rm H}$ within the molecular cloud following a power law whose index is dependent on the propagation mechanism, which itself depends on the turbulence and ion density along the cloud (see, e.g., recent review by \citealp{Padovani2020}). 

A strong grid of correlations is also observed between CS, SO, $^{34}$SO, H$_2$S, and OCS, that is, all the S-bearing molecules with the exception of HCS$^+$. The chemistry leading to CS and HCS$^+$ is thought to be initiated by reactions of S$^+$ and neutral compounds in the external layers of the cloud. However, in dark molecular clouds the formation of SO and OCS is initiated by reactions of atomic S with O$_2$ and OH, while CS is partly formed by the reaction C + SO $\rightarrow$ CS + O \citep{Bulut2020}. The decrease of CS, SO, $^{34}$SO, and OCS with density is more likely related with the adsorption of  sulfur atoms on the grains surface \citep{Vidal2017, Laas2019, Navarro2020}. In order to estimate how sulfur depletion changes with density, we added the abundances of CS, SO, H$_2$S, and OCS, and found that X(S-bearing) decreases with density following the power law X(S-bearing) $\propto$ n$^{-0.6\pm0.1}$. As the abundances of these species correlate almost linearly with the gas-phase elemental sulfur abundance (see e.g., \citealp{Vidal2017}), this would imply that sulfur depletion from the gas phase increases by a factor of $\sim$100 in the n(H$_2$) = 10$^3-10^6$ cm$^{-3}$ range.

\begin{figure*}
\centering
\includegraphics[scale=.50]{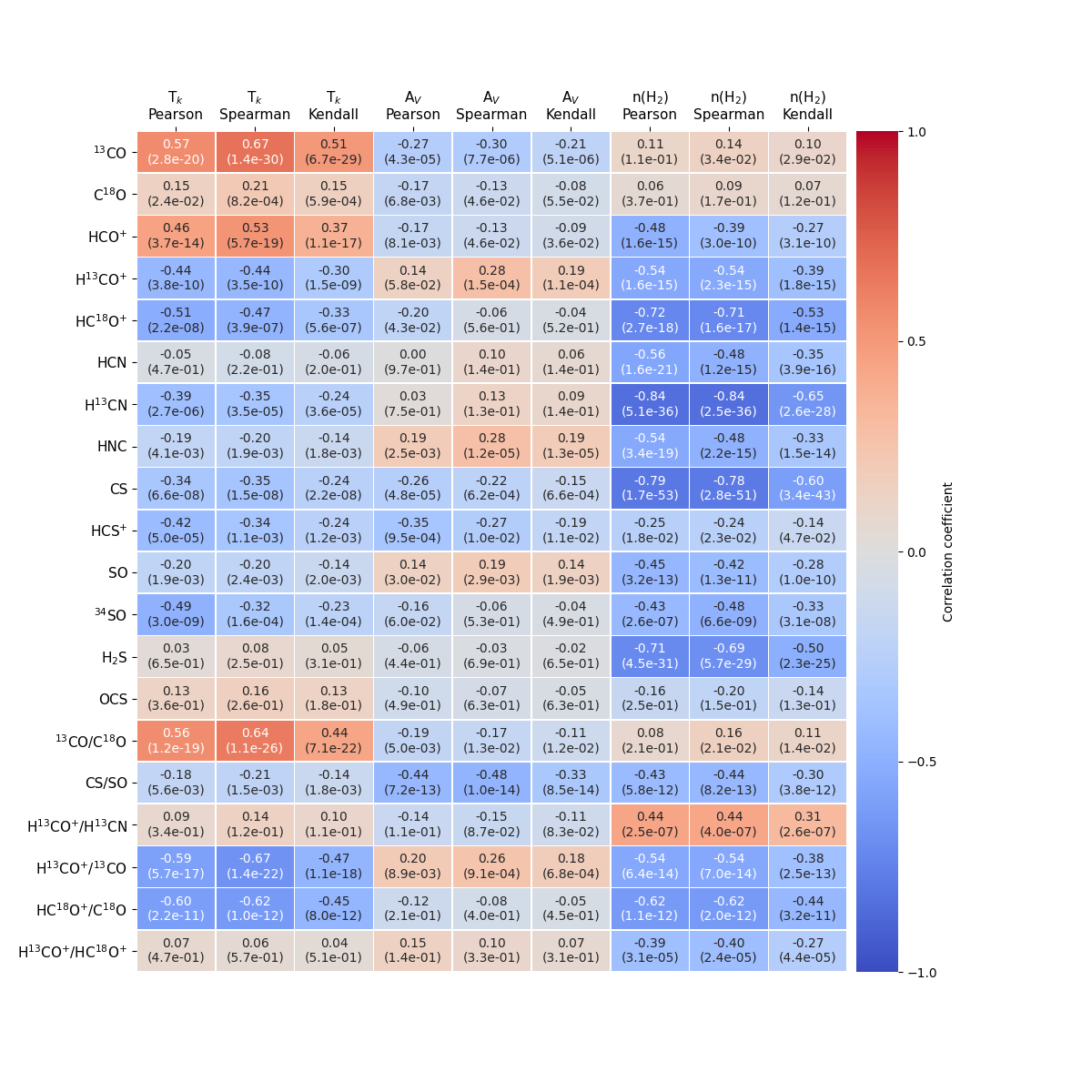}
\caption{Pearson, Spearman, and Kendall correlation parameters of the relations between the molecular abundances of the studied species  (and their ratios) and the molecular cloud physical parameters: kinetic temperature, extinction, and molecular hydrogen density. Numbers in brackets correspond to the null hypothesis probability in each case, i.e. probability of no relation between variables, where a value higher than 0.001 is considered to indicate no ralation at all. }
\label{Fig: Coef_param}
\end{figure*}

\subsection{Relation to T$_k$ and A$_V$}
\label{Subsec: Relation with Tk and Av}

The Pearson, Spearman, and Kendall correlation coefficients for the relations of molecular abundances and a set of selected abundance ratios with the cloud physical parameters (kinetic temperature, extinction, and molecular hydrogen density) are included in Fig. \ref{Fig: Coef_param}.  
The three first columns of Fig. \ref{Fig: Coef_param} show the degree of correlation existing between the molecular abundances and the gas kinetic temperature. Correlation coefficients indicate a positive relation between the molecular abundances of $^{13}$CO and HCO$^{+}$ and T$_k$. The Spearman coefficient is larger than the Pearson one in both cases, pointing to a nonlinear relationship between parameters. This possible correlation becomes weaker for C$^{18}$O and completely disappears for H$^{13}$CO$^+$ and HC$^{18}$O$^+$. This suggests that the correlation of the abundance of $^{13}$CO and HCO$^+$ with T$_k$ is not real, but it is an artifact produced by the large opacity of the observed lines in most positions. This could also partially explain the good correlation of X($^{13}$CO)/X(C$^{18}$O) with gas kinetic temperature. However, it is known that this abundance ratio increases with the local UV field, which is known to be correlated with the dust temperature (see e.g., \citealp{Bron2018, Fuente2019}). This is corroborated by our observations, from which we obtain that the X($^{13}$CO)/X(C$^{18}$O) ratio is higher in Orion than in Perseus, and is higher in Perseus than in Taurus, when comparing positions located at the same visual extinction, even for A$_V <$ 10 mag (see central panel of Fig.~\ref{Fig: Molecular_abundances_rat13COC18O}).

For several other molecules (H$^{13}$CO$^+$, HC$^{18}$O$^+$, H$^{13}$CN, HNC, CS, HCS$^+$, SO, $^{34}$SO) the correlation coefficients in Fig. \ref{Fig: Coef_param} indicate the existence of a weak anti-correlation between molecular abundance and T$_k$.  Confirmation of this trend requires a more detailed analysis. Finally, we do detect significant anti-correlation between H$^{13}$CO$^+$/$^{13}$CO and HC$^{18}$O$^+$/C$^{18}$O and the gas temperature. This effect is also shown in Fig.~\ref{Fig: Molecular_abundances_HCOandfriends}, where the positions belonging to Orion present lower H$^{13}$CO$^+$ and HC$^{18}$O$^+$ abundances than those in Perseus, and the positions
belonging to Perseus present lower H$^{13}$CO$^+$ and HC$^{18}$O$^+$ abundances
than those in Taurus. Indeed, this anti-correlation reflects the different mean abundances of these ions in Taurus, Perseus, and Orion.

Figure \ref{Fig: Coef_param} reveals that there are no clear relations between the complete datasets of molecular abundances and the extinction. The specific analysis of some molecules in Sect.~5 shows that extinction has indeed an influence in some particular aspects of the molecular abundance behavior, as is seen in the case of SO, $^{34}$SO (Sect. \ref{Subsec: SO,34SO}), and mainly $^{13}$CO, C$^{18}$O (Sect. \ref{Subsec: 13CO,C18O}). However, this influence does not follow a monotonic law. We would like to recall the dependence of the $^{13}$CO and C$^{18}$O abundances on visual extinction in Taurus. The correlation is positive for low visual extinctions and negative for high visual extinctions (see Fig.~\ref{Fig: Molecular_abundances_13CO_C18O}), thus preventing a clean result when considering the whole dataset.

\subsection{Relation to n(H$_2$)}
\label{Subsec: Relation with nH2}

The qualitative analysis of the relations between the molecular abundances of the studied species and the gas physical parameters carried out in Sect. \ref{Sec: Molecular column densities} already indicated that the molecular hydrogen density is the main driver of the abundance evolution within molecular clouds. This is confirmed by the correlation coefficients in Fig. \ref{Fig: Coef_param}, pointing to clear anti-correlations between most of the molecular abundances and the molecular hydrogen density. Pearson coefficients are greater than or similar to 0.5 in absolute value for HCO$^+$, H$^{13}$CO$^+$, HC$^{18}$O$^+$, H$^{13}$CN, HNC, CS, H$_2$S, SO, and $^{34}$SO, and the anti-correlation is particularly strong for HC$^{18}$O$^+$, H$^{13}$CN, CS, and H$_2$S, with Pearson coefficients greater than or similar to 0.7. The extremely low values of the probability of the null hypothesis obtained for all these coefficients (at least smaller than 1.0$\times$10$^{-6}$) testify to the reliability of our results. The correlation stands for densities ranging from n(H$_2$)$\sim$10$^3$~cm$^{-3}$ to $\sim$10$^6$~cm$^{-3}$, i.e., over three orders of magnitude.

\begin{figure*}
\centering
\includegraphics[scale=.45]{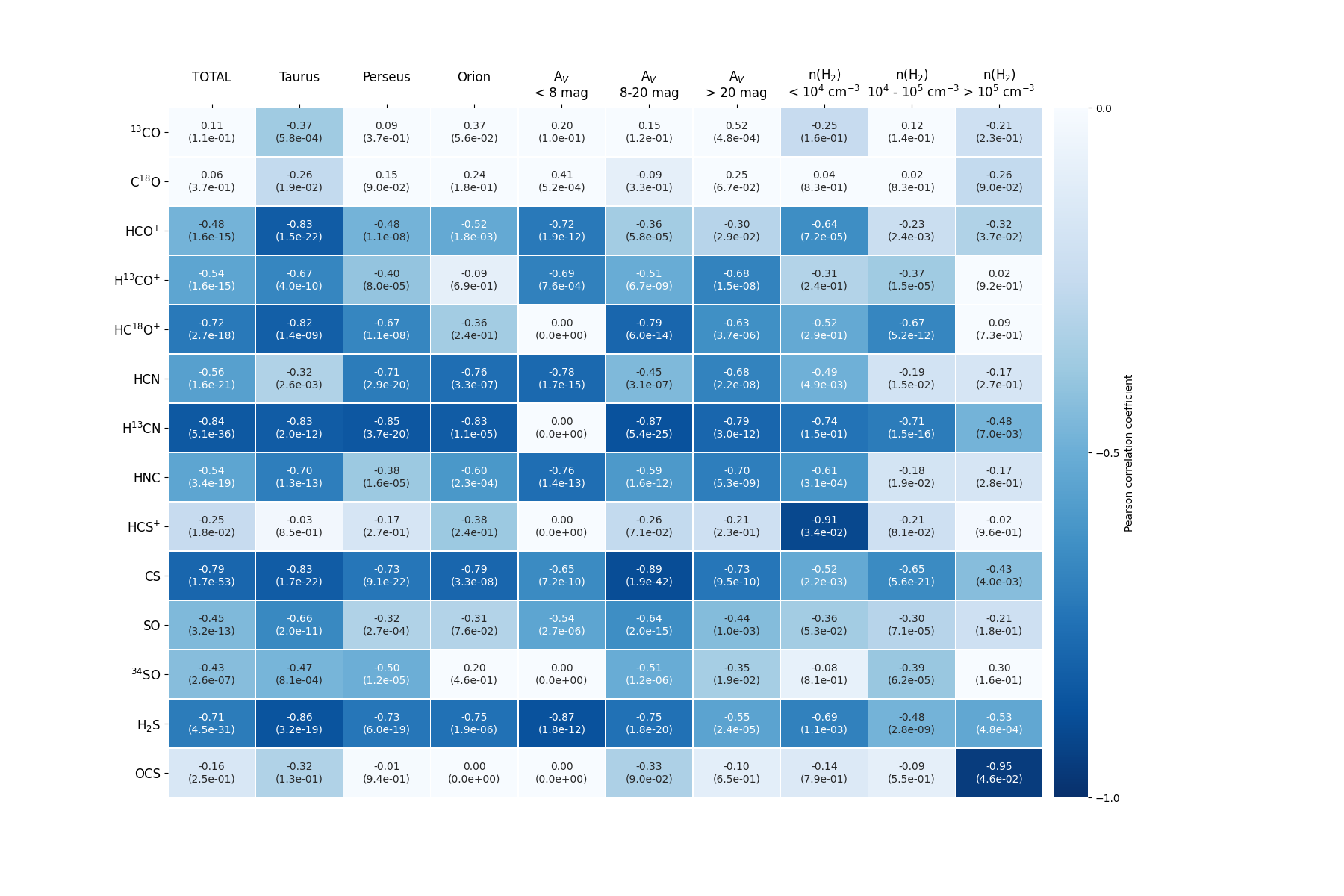}
\caption{Pearson correlation coefficients of the relations between molecular abundances and molecular hydrogen density, considering the total sample (first column), the molecular cloud of the points (second, third, and fourth columns), the extinction bins (fifth, sixth, and seventh columns) and the density bins (eighth, ninth, and tenth columns). Numbers in brackets correspond to the null hypothesis probability in each case, i.e. probability of no relation between variables, where a value higher than 0.001 is considered to indicate no ralation at all.}
\label{Fig: Coef_dens}
%\vspace{-0.1cm}
\end{figure*}

In order to gain deeper insight into these relations, we explored the possible dependence of these anti-correlations on visual extinction and star formation activity. The Pearson coefficients of the relations between molecular abundances and n(H$_2$) considering each one of the clouds, as well as dividing the whole sample applying extinction bins of A$_V$ < 8, 8$-$20 mag, and > 20 mag, are included in Fig. \ref{Fig: Coef_dens}. It can be observed that the correlation with density is always stronger in the case of the Taurus cloud. This is explained by the low incident UV field, turbulence, and gas temperature prevailing in Taurus, where the chemistry is driven
by the freeze-out of molecules on the grain surfaces and $\zeta (H_2)$. Perseus and Orion show weaker anti-correlations with H$_2$ density, with smaller correlation parameters. However, we do recover strong anti-correlations with density for bins with 8$-$20 mag in Perseus and $>$20 mag in Orion. In these deep layers of the Perseus and Orion molecular clouds the gas chemistry is determined by the molecular hydrogen density, as in Taurus. This is a chemical footprint of the existence of a dark region in these clouds, where the cosmic ray flux and the molecular depletion are driving the chemistry as in Taurus. This dark region is located at a different visual extinction depending on the star formation activity in the neighborhood. In particular, our data show that this dark region is located at A$_V >$ 8-10 mag in Perseus and at A$_V >$ 20 mag in Orion.

\section{Limitations of our study}
\label{Subsec: error}

We were not able to carry out a multi-transition study of every species. For some species, we only observed one line, precluding any molecular excitation study (see Table B.2 in \citealp{Fuente2019}). Even for the species for which we observed several transitions, such as SO and H$_2$CS, we only detected one transition in many positions. In the case of H$_2$CS, its lines have only been detected in a few positions (those close to the extinction peaks) and we did not consider this species here. In order to calculate molecular column densities and abundances in a uniform way and in a large number of positions, we calculated molecular abundances assuming the density derived from the multi-transition study of CS. The sensitivity of the estimated molecular abundances to the assumed gas density depends on the critical density of the transition used in the calculation. For densities higher than the critical density, T$_{ex}\approx$T$_k$, and the molecular column density no longer depends on the density. This is the case for the $^{13}$CO and C$^{18}$O J=1$\rightarrow$0 lines, for which the estimated column densities are not affected by possible uncertainties in the adopted densities. However, for most of the studied lines the critical density is higher than those measured in our positions (see Table~\ref{Table: Critical densities}). Under these conditions, the lines are subthermally excited and T$_{ex}\propto$ n(H$_2$)$\times$ N$_X$. An error in the molecular hydrogen density therefore translates to an error of approximately the same order in the molecular column density.  

To further test the reliability of the assumed densities, we independently estimated the gas densities in 28 positions from a multi-transition study of SO and H$_2$CS. These positions correspond to high visual extinction peaks where several SO and H$_2$CS lines are detected. Our results show that the densities derived from CS (and C$^{34}$S, $^{13}$CS), SO ($^{34}$SO), and H$_2$CS are in agreement within a factor of approximately five. Therefore, we consider that column densities towards the positions with n(H$_2$)$<$n$_{crit}$ are reliable within this same factor. 

An important assumption in our calculations is that the gas along the line of sight presents uniform physical and chemical conditions. This assumption is reasonable at the layers with A$_V<$8$-$10 mag, where the gas kinetic temperature and density gradients are shallow, but it might not apply to the positions close to the extinction peaks. In order to discuss the possible impact of this assumption, we consider a toy model with two phases, "Phase 1" corresponding to a dense core with n(H$_2$)=10$^5$ cm$^{-3}$, and "Phase 2" corresponding to a moderate density envelope with n(H$_2$)=10$^4$ cm$^{-3}$. In this simple model, the observed brightness temperature would be

\begin{equation}
T_b=  T_1 \times exp(-\tau_2)+  J_\nu(T_{ex,2})  \times (1-exp(-\tau_2))
,\end{equation}
\\
where T$_1$ is $ T_1 = J_\nu(T_{ex,1}) \times (1-exp(-\tau_1))$, with $T_{ex,1}$ and $\tau_1$ being the excitation temperature and line opacity of the dense phase. Similarly, $T_{ex,2}$ and $\tau_2$ are the excitation temperature and line opacity of "Phase 2". If we have high values of $\tau_2$ and low values of T$_{ex,2}$, which is the case for the millimeter lines of molecules with a high dipole moment, the flux emergent from the cloud would be $\sim$ T$_1$ $\times$ exp($-\tau_2$), producing self-absorbed profiles. This kind of self-absorbed profile is observed in the J=1$\rightarrow$0 lines of the abundant isotopologs HCO$^+$, HCN, and HNC (see examples of spectra in \citealp{Fuente2019}). In the case of the rarer isotopologs H$^{13}$CO$^+$, HC$^{18}$O$^+$, H$^{13}$CN, C$^{34}$S, $^{13}$CS, and $^{34}$SO, the lines are expected to be optically thin, and T$_b$= T$_{ex,1}$ $\times \tau_1$ + T$_{ex,2}$ $\times \tau_2$, which is proportional to the sum of the column densities weighted by the excitation temperatures. In this case, our one-phase model is giving an averaged excitation temperature and column density along the line of sight. This approximation is reasonable as long as the molecular abundances do not greatly differ between the two phases. In the most pessimistic case of one species that is only abundant in one phase, the derived molecular abundance would be wrong by a factor of ten, which is the difference between the Phase 1 and 2 densities. All the studied species are abundant in the translucent part of the cloud, as demonstrated in TMC~1 by \citet{Fuente2019}. Although different species and transitions are probing different regions within the cloud, the observed transitions have similar E$_u$ and n$_{crit}$ to those of CS, and therefore are expected to come from the same gas layer. 

To further check our result, in Fig.~\ref{Fig: Coef_dens} we show the relations between density and molecular abundances in different bins of visual extinction and density. The similarity in the correlation coefficients found for all the bins supports the reliability of our results.

\begin{table}
\caption{Critical densities}
\label{Table: Critical densities}
\centering
\resizebox{\columnwidth}{!}{% 
\begin{tabular}{llccc}\\%l}\\
%\multicolumn{3}{l}{Table 1. GEMS sample} \\ 
\hline\hline
\noalign{\smallskip}
Molecule & Transition  & A$_{ij}$ &  C$_{ij}$  & n$_{crit}$ \\%& CD(CS)\\
              &                    &(s$^{-1}$) & (T$_k$=10~K) & (cm$^{-3}$)\\
\hline
\noalign{\smallskip}                          
CS                      &      2$\rightarrow$1         &   1.679$\times$10$^{-5}$    &    4.7$\times$10$^{-11}$   &   3.3$\times$10$^5$     \\
C$^{18}$O        &  1$\rightarrow$0              &    7.203$\times$10$^{-8}$   &     3.3$\times$10$^{-11}$   &   2.0$\times$10$^3$     \\
H$^{13}$CO$^+$   &   1$\rightarrow$0        &   3.853$\times$10$^{-5}$    &    2.6$\times$10$^{-10}$   &   1.4$\times$10$^5$    \\
HCS$^+$                &    1$\rightarrow$0      &   1.109$\times$10$^{-5}$    &    3.8$\times$10$^{-10}$    &   2.9$\times$10$^4$     \\
H$^{13}$CN      &    1$\rightarrow$0          &   2.225$\times$10$^{-5}$   &    3.0$\times$10$^{-11}$     &   7.2$\times$10$^5$   \\
SO                    &      3$_4$$\rightarrow$2$_3$      &   3.165$\times$10$^{-5}$      &  8.3$\times$10$^{-11}$    &  3.7$\times$10$^5$  \\
HNC                  &    1$\rightarrow$0                       &   2.690$\times$10$^{-5}$      &  9.7$\times$10$^{-11}$     &  2.8$\times$10$^5$  \\
OCS                  &    7$\rightarrow$6                       &   1.714$\times$10$^{-6}$      &  7.6$\times$10$^{-11}$     &  2.2$\times$10$^4$    \\
H$_{2}$S          & 1$_{10}$$\rightarrow$1$_{01}$  &   2.653$\times$10$^{-5}$     &  3.4$\times$10$^{-11}$     &   4.8$\times$10$^5$  \\
\noalign{\smallskip}
\hline \hline

\end{tabular}}
\end{table}

\section{Comparison between Taurus, Perseus, and Orion}

The C$^{18}$O column density is widely used as a tracer of the gas mass in our galaxy and other galaxies. The reliability of this tracer is based on presenting a relatively uniform fractional abundance in interstellar molecular clouds. Based on our data, we estimated
the mean C$^{18}$O abundances in Taurus, Perseus, and Orion. The obtained values are:  X(C$^{18}$O) = (1.4 $\pm$ 0.7) $\times$ 10$^{-7}$ for Taurus, X(C$^{18}$O) = (2.1 $\pm$ 1.8) $\times$ 10$^{-7}$ for Perseus, and X(C$^{18}$O) = (1.6 $\pm$ 1.0) $\times$ 10$^{-7}$ for Orion. These values are in good agreement with those obtained by previous works, as \citet{Frerking1989} in Taurus, \citet{Trevino2019} in Monoceros R2, and  \citet{Roueff2020} in Orion B. Therefore, our results corroborate the idea that the abundance of C$^{18}$O is quite stable at scales of molecular clouds, and is therefore a good gas mass tracer for large-scale surveys and extragalactic research.

We derived the mean values of the $^{13}$CO/C$^{18}$O abundance ratio for the three star-forming regions, obtaining values of $^{13}$CO/C$^{18}$O = 10.9$\pm$7.5 for Taurus, $^{13}$CO/C$^{18}$O = 17.3$\pm$11.5 for Perseus, and $^{13}$CO/C$^{18}$O = 23.7$\pm$10.4 for Orion. The expected value is $^{13}$CO/C$^{18}$O = 7.5 $-$ 9.8, assuming $^{12}$C/$^{13}$C = 57 $-$ 67 and $^{16}$O/$^{18}$O = 500 $-$ 600 \citep{Gerin2015, Langer1990, Wilson1994}. As mentioned in Sect. \ref{Subsec: 13CO,C18O}, the $^{13}$CO/C$^{18}$O abundance ratio increases in regions of enhanced UV field \citep{Shimajiri2014, Ishii2019, Areal2018}. This variation is interpreted in terms of the selective photodissociation and isotopic fractionation (see \citealp{Bron2018, Fuente2019}). This effect is produced because the more abundant CO isotopolog shields itself from the effect of UV photons more efficiently than less abundant isotopologs \citep[][]{Stark2014, Visser2009}. Our data confirm this trend in Orion, although the observed positions are located at a distance $>$ 1 pc from the ionized nebula M~42, which shows that the whole cloud is illuminated by a strong UV field emitted by young massive stars in the Trapezium cluster \citep{Pabst2019}. In contrast to X(C$^{18}$O), which presents uniform abundance in the three regions, the mean $^{13}$CO abundance is a factor of approximately two higher in Orion than in Taurus. One interesting question pertains to whether the observed variation in the $^{13}$CO abundance is related to a variation in the $^{12}$CO/$^{13}$CO ratio \citep{Roueff2015, Colzi2020} or is revealing variations for a higher $^{12}$CO abundance in Orion. Because of the higher gas temperature and incident UV field in Orion, selective photodissociation would work in the direction of increasing N($^{12}$CO)/N($^{13}$CO) in Orion with respect to Taurus. Therefore, a higher $^{12}$CO abundance in Orion stands as the most likely explanation. Our results are based on significant approximations (one single phase, gas-dust thermalization), and therefore a detailed multi-transition study of these compounds is required to confirm this result. 

Apart from $^{13}$CO, C$^{18}$O, and H$_2$S, the abundances of all the studied species decrease following the sequence Taurus, Perseus, and Orion. This might be interpreted considering the observed relations between the molecular abundances and the molecular hydrogen density, and the different density distributions of the positions observed in Taurus, Perseus, and Orion (see Fig.~\ref{Fig: Histograms_dens_pres}). It should be noted that the relation between CS abundance and density does not present significant differences between molecular clouds. However, the abundances of the protonated compounds HCO$^+$ (and isotopologs) and HCS$^+$ are systematically lower in Orion than in Taurus and Perseus for similar values of the molecular hydrogen density and visual extinction. To a lesser extent, this behavior is also observed in H$^{13}$CN. A detailed modeling of the chemistry including all the species is required to discern the cause of this difference.

\section{The extragalactic connection}
\label{The extra-galactic connection}

Our data provide a comprehensive view of the abundance behavior of the most commonly observed species in molecular clouds, and  it is interesting to compare them with extragalactic chemical studies. In most cases, the spatial resolution of current telescopes does not allow molecular clouds in external galaxies to be resolved, and therefore millimeter observations are used to measure cloud-weighted molecular abundances. To compare with external galaxies we computed mean values of observed molecular column densities and abundances in Taurus, Perseus, and Orion (Table \ref{Table: Xgal}). We caution that these mean values were computed using only detections. The majority of the detections correspond to dense gas (n(H$_2$)$>$10$^4$ cm$^{-3}$) located in regions with A$_V>$8 mag and therefore the obtained abundances are representative of the dense phase. We would also like to reiterate the selection criteria applied for the positions of the GEMS program. The positions were selected to avoid protostars and shocks. Nevertheless, there are differences between the abundances measured in Taurus, Perseus, and Orion. The greatest differences between these averaged values are found for the N($^{13}$CO)/N(C$^{18}$O) ($R^{13CO}_{C18O}$) and the N(HCO$^+$)/N(H$^{13}$CO$^+$) ($R^{HCO+}_{HC13O+}$) ratios. Both $R^{13CO}_{C18O}$ and $R^{HCO+}_{HC13O+}$ increase with star formation activity following the sequence Taurus, Perseus, and Orion. This progressive increase is related to the different density distribution, average gas and dust temperatures, and incident UV field. 
 
In Table \ref{Table: Xgal}, we compare the mean column densities in Taurus, Perseus, and Orion with those observed in a sample of galaxies including the starburst galaxies M 83, M 82, and NGC 253, the galaxies hosting an active galactic nucleus (AGN) M 51, NGC 1068, and NGC 7469, and the ultra-luminous infrared galaxies (ULIRGs) Arp 220 and Mrk 231. \citet{Aladro2015} carried out a molecular survey of this sample at $\lambda$=3~mm using the IRAM 30m telescope, producing a uniform database that can be easily compared with our observations. Moreover, the molecular abundances were derived using the same rotational lines as in this work. \citet{Aladro2015} did not detect H$^{13}$CO$^+$ towards NGC~1068. We used the data reported by \citet{Usero2004} to estimate N(HCO$^+$) and N(H$^{13}$CO$^+$) by assuming the same source size and using the same methodology as \citet{Aladro2015} (LTE and T$_{rot}$=10~K). The obtained values are shown in Table \ref{Table: Xgal}.

The values of  $R^{13CO}_{C18O}$ are lower in the sample of galaxies considered than in GEMS sample. Moreover, we do not observe any increase with star formation activity. Instead, N($^{13}$CO)/N(C$^{18}$O)$\sim$3$-$6 in starbursts and AGNs, and decreases to $\sim$1 in ULIRGs. This suggests that at scales of the whole galaxy, $R^{13CO}_{C18O}$ is not determined by the star formation activity but by optical depth effects and variations in the carbon and oxygen isotopic ratio because of stellar nucleosynthesis \citep{Henkel1993, Henkel1998, Wang2004}. \citet{Donaire2017} carried out a survey of $^{13}$CO and C$^{18}$O in a sample of nine nearby spiral galaxies. These latter authors found an average value of
$R^{13CO}_{C18O}$=7.9$\pm$0.8 measured in the whole galaxy and $R^{13CO}_{C18O}$=6.0$\pm$0.9 for the central regions of the galaxies, that is, $R^{13CO}_{C18O}$ seems to be anti-correlated with star formation activity. These same authors, using higher spatial resolution ALMA observations (beam$\sim$8$" <$ 400 pc) of  NGC 3351, NGC 3627, and NGC 4321, observed that when zooming into the central regions, $R^{13CO}_{C18O}$ increases in a narrow ring at R$\sim$500 pc. As the authors discussed, this increase in $R^{13CO}_{C18O}$ might be related to the enhanced star formation in these regions, which are fed by gas flows along the bars. Thus, the $R^{13CO}_{C18O}$ could be a diagnostic of star formation at scales comparable to the size of a giant molecular cloud, namely approximately a few hundred parsecs, but fails at scales of the whole galaxy.  Moreover, accurate determination of the local $^{12}$C/$^{13}$C and $^{16}$O/$^{18}$O isotopic ratios is required for the correct comparison with galactic patterns and chemical models. 

In the GEMS sample, the N(HCO$^+$)/N(H$^{13}$CO$^+$) ratio is correlated with local star formation activity. Despite the limitations of our comparison, we observe the same 
trend in the galaxies sample. In M~83,  $R^{HCO+}_{HC13O+}$ is similar to those measured in Taurus and Perseus, suggesting that the J=1$\rightarrow$0 emission of these species is dominated by low- and intermediate-mass star forming regions. On the contrary, the other galaxies present $R^{HCO+}_{HC13O+}$ $>$10, suggesting that the emission could be dominated by massive star forming regions. In Taurus, the low estimated value of the  N(HCO$^+$)/N(H$^{13}$CO$^+$) ratio is not a chemical effect, but the consequence of a thick absorbing envelope that leads to a systematic underestimation of the HCO$^+$ abundance when using  one-phase models and the optically thick J=1$\rightarrow$0 line.  This thick envelope seems to be absent in Orion more likely because of the enhanced UV field that photodissociates the molecules at low visual extinctions. The similarity between the N(HCO$^+$)/N(H$^{13}$CO$^+$) in Orion and those in starburst might be the consequence of the cloud properties in these regions where the ambient UV field is expected to be higher than in our galaxy (see, e.g.,  \citealp{Fuente2005, Fuente2006, Fuente2008}). The same trend is observed in N(HCN/N(H$^{13}$CN).

It is interesting to discuss the N(H$^{13}$CO$^+$)/N(H$^{13}$CN) (or N(HCO$^+)$)/N(HCN)) ratio, which is commonly used to differentiate between starburst- and AGN-dominated chemistry (see, e.g., \citealp{GBurillo2008}). Indeed, the abundance of H$^{13}$CO$^+$ is highest in the M~82 galaxy which is considered as the prototype of evolved  starburst with a photon-dominated chemistry in the nuclear region \citep{Fuente2005, Fuente2006, Fuente2008}. In contrast, H$^{13}$CN is especially abundant in the active galaxy NGC 1068 and the ULIRGS Arp 220 and Mrk 231. The extraordinary abundance of HCN in these galaxies have been widely discussed by several authors that proposed mechanical heating, shocks, and X-rays as physical agents to boost the abundance of HCN \citep{Usero2004, GBurillo2008, Perez2009, GBurillo2014, Viti2014}. We do not see a clear variation of the  N(H$^{13}$CO$^+$)/N(H$^{13}$CN) in the GEMS sample. We recall that GEMS positions were selected to avoid protostars, HII regions, and bipolar outflows. The difference between the N(H$^{13}$CO$^+$)/N(H$^{13}$CN) ratios in GEMS clouds and those towards AGNs and ULIRGS supports the interpretation that the emission of H$^{13}$CN in these galaxies is dominated by the gas associated with XDRs and shocks.

\citet{Viti2017} analyzed grids of time-dependent chemical models, varying in gas density, temperature, cosmic ray ionization rate, and radiation field to calculate abundances and line intensities in order to compare with molecular observations in external galaxies. This latter study found that line intensities and line ratios from different chemical models can be very similar, leading to a large degeneracy. This degeneracy can be partially removed if chemical abundances and abundance ratios are used instead for the comparison. We provide tables of chemical abundances for three Galactic prototypical star forming regions that could be useful to interpret molecular data in external galaxies even if our data are not representative of the whole molecular cloud.

\begin{table*}
\caption{Mean column densities obtained in this work for Taurus, Perseus, and Orion and column densities observed in a sample of nearby galaxies. The galaxy sample data are obtained from \citet{Aladro2015} (upper limits are shown for undetected species), with the exception of the HCO$^+$ and H$^{13}$CO$^+$ column densities in  NGC\,1068, obtained from \citet{Usero2004}. Star formation rates of the comparison galaxies are obtained from \citet{Walter2008} for M\,83; \citet{Strickland2004} for NGC\,253 and M\,82; \citet{Schuster2007} for M\,51; \citet{Esquej2014} for NGC\,1068; \citet{Genzel1995} for NGC\,7469; \citet{Anantharamaiah2000} for Arp\,220; \citet{Taylor1999} for Mrk\,231. This parameter is only meant to give a rough idea of the activity, as it was calculated for different volumes in each galaxy.  Star formation rate in galactic molecular clouds are taken from \citet{Lada2010}.}
\label{Table: Xgal}
\centering     
\resizebox{\textwidth}{!}{%    
\begin{tabular}{llllllllllll}\\%l}\\
%\multicolumn{8}{l}{Table 1. GEMS sample} \\ 
\hline\hline
\noalign{\smallskip}
Molecule        &       Taurus  &       Perseus &       Orion   &       M\,83   &       NGC\,253        &       M\,82   &       M\,51   &       NGC\,1068       &       NGC\,7469       &       Arp\,220        &       Mrk\,231\\%& CD(CS)\\
%& & ID & RA (J2000) & Dec (J2000) & & & & (cm$^{-3}$) \\%& (cm$^{-2}$)\\
%\hline
\noalign{\smallskip}
\hline
\noalign{\smallskip}
SFR     (M$_\odot$ yr$^{-1}$) &   715$\cdot10^{-6}$ &   150$\cdot10^{-6}$       &  715$\cdot10^{-6}$      &       2.5     &       3.6     &       10      &       2.5     &       0.4     &       30      &       240     &       220 
\\                 
$^{13}$CO       &       1.1$\cdot10^{16}$       &       3.8$\cdot10^{16}$       &       7.5$\cdot10^{16}$       &       5.7$\cdot10^{16}$       &       1.9$\cdot10^{17}$       &       1.3$\cdot10^{17}$       &       2.1$\cdot10^{16}$       &       4.6$\cdot10^{17}$       &       3.6$\cdot10^{16}$       &       6.6$\cdot10^{17}$       &       5.6$\cdot10^{16}$       \\

C$^{18}$O       &       1.4$\cdot10^{15}$       &       3.0$\cdot10^{15}$       &       4.6$\cdot10^{15}$       &       1.1$\cdot10^{16}$       &       5.1$\cdot10^{16}$       &       2.6$\cdot10^{16}$       &       5.2$\cdot10^{15}$       &       1.3$\cdot10^{17}$       &       6.4$\cdot10^{15}$       &       6.2$\cdot10^{17}$       &       4.9$\cdot10^{16}$       \\

HCO$^+$ &       3.1$\cdot10^{12}$       &       6.3$\cdot10^{12}$       &       2.6$\cdot10^{13}$       &       4.6$\cdot10^{13}$       &       1.9$\cdot10^{14}$       &       2.6$\cdot10^{14}$       &       1.3$\cdot10^{13}$       &       5.5$\cdot10^{14}$       &       7.4$\cdot10^{13}$       &       1.1$\cdot10^{15}$       &       3.3$\cdot10^{14}$       \\

H$^{13}$CO$^+$  &       1.7$\cdot10^{12}$       &       1.5$\cdot10^{12}$       &       1.6$\cdot10^{12}$       &       1.1$\cdot10^{13}$       &       1.3$\cdot10^{13}$       &       6.9$\cdot10^{12}$       &       $\leq$6.6$\cdot10^{11}$ &       2.0$\cdot10^{13}$       &       $\leq$8.0$\cdot10^{12}$ &       9.8$\cdot10^{13}$       &       $\leq$9.1$\cdot10^{13}$ \\

HC$^{18}$O$^+$  &       1.4$\cdot10^{11}$       &       1.3$\cdot10^{11}$       &       1.4$\cdot10^{11}$       &       0       &       4.3$\cdot10^{12}$       &       4.70$\cdot10^{12}$      &       0       &       0       &       0       &       0       &       0       \\

HCN     &       4.8$\cdot10^{13}$       &       5.5$\cdot10^{13}$       &       2.2$\cdot10^{14}$         &       8.6$\cdot10^{13}$       &       3.8$\cdot10^{14}$       &       2.9$\cdot10^{14}$         &       4.5$\cdot10^{13}$       &       2.4$\cdot10^{15}$       &       1.0$\cdot10^{14}$         &       4.6$\cdot10^{15}$       &       8.5$\cdot10^{14}$ \\

H$^{13}$CN      &       4.7$\cdot10^{12}$       &       4.5$\cdot10^{12}$       &       4.9$\cdot10^{12}$       &       2.3$\cdot10^{12}$       &       2.8$\cdot10^{13}$       &       5.0$\cdot10^{12}$       &       1.8$\cdot10^{12}$       &       8.1$\cdot10^{13}$       &       $\leq$1.4$\cdot10^{13}$ &       7.1$\cdot10^{14}$       &       2.0$\cdot10^{14}$       \\

HNC     &       1.3$\cdot10^{13}$       &       1.8$\cdot10^{13}$       &       5.7$\cdot10^{13}$       &       3.1$\cdot10^{13}$       &       1.8$\cdot10^{14}$       &       1.2$\cdot10^{14}$       &       1.4$\cdot10^{13}$       &       7.0$\cdot10^{14}$       &       4.2$\cdot10^{13}$       &       3.2$\cdot10^{15}$       &       2.5$\cdot10^{14}$       \\

HCS$^+$ &       4.5$\cdot10^{11}$       &       4.3$\cdot10^{11}$       &       3.5$\cdot10^{11}$       &       0       &       0       &       0       &       0       &       0       &       0       &       0       &       0       \\

CS      &       6.1$\cdot10^{13}$       &       5.6$\cdot10^{13}$       &       5.5$\cdot10^{13}$       &       5.7$\cdot10^{13}$       &       4.1$\cdot10^{14}$       &       2.5$\cdot10^{14}$       &       1.8$\cdot10^{13}$       &       7.6$\cdot10^{14}$       &       8.1$\cdot10^{13}$       &       3.4$\cdot10^{15}$       &       3.5$\cdot10^{14}$       \\

SO      &       2.0$\cdot10^{13}$       &       3.8$\cdot10^{13}$       &       2.1$\cdot10^{13}$       &       2.0$\cdot10^{13}$       &       1.7$\cdot10^{14}$       &       7.7$\cdot10^{13}$       &       1.1$\cdot10^{13}$       &       2.2$\cdot10^{14}$       &       $\leq$6.9$\cdot10^{13}$ &       $\leq$4.4$\cdot10^{14}$ &       $\leq$5.0$\cdot10^{14}$ \\

$^{34}$SO       &       1.8$\cdot10^{12}$       &       3.3$\cdot10^{12}$       &       7.6$\cdot10^{11}$       &       0       &       0       &       0       &       0       &       0       &       0       &       0       &       0       \\

H$_2$S  &       2.4$\cdot10^{13}$       &       5.0$\cdot10^{13}$       &       7.2$\cdot10^{13}$       &       0       &       0       &       0       &       0       &       0       &       0       &       0       &       0       \\

OCS     &       5.2$\cdot10^{12}$       &       1.1$\cdot10^{13}$       &       0       &       $\leq$7.3$\cdot10^{13}$ &       4.9$\cdot10^{14}$       &       $\leq$8.4$\cdot10^{13}$ &       $\leq$7.4$\cdot10^{13}$ &       $\leq$8.9$\cdot10^{14}$ &       $\leq$8.2$\cdot10^{14}$ &       $\leq$6.1$\cdot10^{15}$ &       $\leq$9.2$\cdot10^{15}$ \\

$^{13}$CO/C$^{18}$O$^{(1)}$     &       10.9    &       17.3    &       23.7    &       5.2     &       3.7     &       5       &       4       &       3.5     &       5.6     &       1       &       1.1     \\

HCO$^+$/H$^{13}$CO$^{+}$\,$^{(1)}$      &       1.8     &       4.3     &       16      &       4.1     &       14.9    &       37      &       >20     &       27      &       >9      &       11      &       >3.6    \\

HCN/H$^{13}$CN$^{(1)}$  &       10.3    &       12.4    &       45.5    &       37.4         &       13.8    &       58      &       25      &       30.3    &       7.4     &       6.5     &       4.3 \\

CS/SO   &       5.4     &       4.1     &       3.2     &       2.8     &       2.4     &       3.2     &       1.6     &       3.4     &       1.2     &       >7      &       >0.7    \\

H$^{13}$CO$^+$/$^{13}$CO        &       1.5$\cdot10^{-4}$       &       3.7$\cdot10^{-5}$       &       8.6$\cdot10^{-6}$       &       1.9$\cdot10^{-4}$       &       6.9$\cdot10^{-5}$       &       5.3$\cdot10^{-5}$       &       <3$\cdot10^{-5}$        &       4.3$\cdot10^{-5}$       &       <0.00022        &       1.40$\cdot10^{-4}$      &       <1.6$\cdot10^{-3}$      \\

HC$^{18}$O$^+$/C$^{18}$O        &       9.8$\cdot10^{-5}$       &       3.6$\cdot10^{-5}$       &       1.9$\cdot10^{-5}$       &               &       8.4$\cdot10^{-5}$       &       1.3$\cdot10^{-4}$       &               &               &               &               &               \\

HCO$^+$/HCN     &       0.14    &       0.19    &       0.24    &       0.53    &       0.50    &       0.88    &       0.29    &       0.37    &       0.71    &       0.24    &       0.39 \\

H$^{13}$CO$^+$/H$^{13}$CN       &       0.69    &       0.68    &       0.83    &               &       0.46    &       1.38    &               &       0.25    &               &       0.14    &       <0.45   \\

X(H$^{13}$CO$^+$)       &       1.3$\cdot10^{-10}$      &       6.4$\cdot10^{-11}$      &       2.9$\cdot10^{-11}$      &       1.6$\cdot10^{-10}$      &       4.8$\cdot10^{-11}$      &       4.2$\cdot10^{-11}$      &       <2$\cdot10^{-11}$       &       2.5$\cdot10^{-11}$      &       <2$\cdot10^{-10}$       &       2.5$\cdot10^{-11}$      &       <3$\cdot10^{-10}$       \\

X(H$^{13}$CN)   &       2.9$\cdot10^{-10}$      &       1.9$\cdot10^{-10}$      &       7.7$\cdot10^{-11}$      &       3.4$\cdot10^{-11}$      &       8.8$\cdot10^{-11}$      &       3.0$\cdot10^{-11}$      &       5.5$\cdot10^{-12}$      &       1.0$\cdot10^{-10}$      &       <3.5$\cdot10^{-10}$     &       8.2$\cdot10^{-10}$      &       6.5$\cdot10^{-10}$      \\

X(CS)   &       6.3$\cdot10^{-9}$       &       3.9$\cdot10^{-9}$       &       2.2$\cdot10^{-9}$       &       8.3$\cdot10^{-9}$       &       1.3$\cdot10^{-9}$       &       1.5$\cdot10^{-9}$       &       5.5$\cdot10^{-11}$      &       9.0$\cdot10^{-10}$      &       2.0$\cdot10^{-9}$       &       8.7$\cdot10^{-10}$      &       1.1$\cdot10^{-9}$       \\

X(SO)   &       1.8$\cdot10^{-9}$       &       2.0$\cdot10^{-9}$       &       7.3$\cdot10^{-10}$      &       2.9$\cdot10^{-10}$      &       5.3$\cdot10^{-10}$      &       4.7$\cdot10^{-10}$      &               &               &               &               &               \\

X($^{34}$SO)    &       1.3$\cdot10^{-10}$      &       1.4$\cdot10^{-10}$      &       1.5$\cdot10^{-11}$      &               &               &               &               &               &               &               &               \\

\noalign{\smallskip}                                      
\hline \hline
\end{tabular}}
$^{(1)}$ We would like to reiterate that the low values of the HCO$^{+}$/H$^{13}$CO$^{+}$ in Taurus and Perseus are mainly due to the proliferation of self-absorbed profiles in the HCO$^{+}$ 1-0 spectra in the observed positions. These values need to be understood as lower limits to the real column density ratios. To a lesser extent, this effect is also affecting the HCN/H$^{13}$CN and $^{13}$CO/C$^{18}$O ratios.
\end{table*}

\section{Summary and conclusions}
\label{Sec: Summary and conclusions}

We present the molecular database of the GEMS project. This program is focused on the observation of starless cores in filaments of the nearby star-forming regions Taurus, Perseus, and Orion. These regions have different degrees of star formation activity, and therefore different physical conditions, providing a possibility to explore the effect of environment on gas chemistry. The project includes observations towards 305 positions distributed in 27 cuts that have been selected to avoid recently formed stars and their associated bipolar outflows. The gas chemistry in these positions is therefore not affected by the shocks produced by these energetic phenomena and/or the UV radiation coming from internal sources. Our project is characterizing the molecular cloud chemistry during the pre-stellar phase. The number of positions allows an unprecedented analysis of the statistical trends shown by the molecular abundances in a wide range of physical conditions. 

We carried out a multi-transition analysis of the CS molecule and its isotopologs C$^{34}$S and $^{13}$CS to derive the gas physical conditions, applying the MCMC methodology with a Bayesian inference approach, and using the RADEX code. We derived the molecular hydrogen abundance for 244 positions. Assuming the molecular hydrogen densities derived from the CS multi-transition fitting and using the RADEX code, we determined the molecular abundances for the following species: $^{13}$CO, C$^{18}$O, HCO$^+$, H$^{13}$CO$^+$, HC$^{18}$O$^+$, H$^{13}$CN, HNC, HCS$^+$, SO, $^{34}$SO, H$_2$S, and OCS. We estimated the o-H$_2$S abundances using the specific collisional rates recently calculated by \citet{Dagdigian2020}. We analyzed the relation between the molecular abundances and the gas physical parameters (kinetic temperature, extinction, and molecular hydrogen density), exploring the degree of correlation between parameters by the computation of the Pearson, Spearman, and Kendall correlation coefficients. Our main results are as follows: 

\begin{itemize}
\item  Taurus shows the lowest mean density, with a peaky distribution at  n(H$_2$)$\sim$2$\times$10$^4$cm$^{-3}$ Higher density values, and wider density distributions are found in Perseus and Orion, with mean values of n(H$_2$)$\sim$4$\times$10$^4$cm$^{-3}$
and $\sim$6$\times$10$^4$cm$^{-3}$, respectively. The wider density distribution in Perseus and Orion is the consequence of
the superposition of regions with different environmental conditions resulting from the feedback of recently formed stars.

\item Relations between molecules themselves reveal strong linear correlations that define three families of species: (i) the CO isotopologs $^{13}$CO, C$^{18}$O; (ii) H$^{13}$CO$^+$, HC$^{18}$O$^+$, H$^{13}$CN, and HNC; and (iii) the S-bearing molecules CS, SO, $^{34}$SO, H$_2$S, and OCS.

\item Only $^{13}$CO and C$^{18}$O show a correlation with gas kinetic temperature. In the shielded gas, we observe an increasing relation between the abundances of these species with the gas kinetic temperature, until T$_K \sim$~15~K. Beyond this value, the abundance remains  constant with a large scatter. The X($^{13}$CO)/X(C$^{18}$O) increases in the cloud border as the consequence of selective photodissociation and isotopic fractionation. 

\item The abundances of H$^{13}$CO$^+$, HC$^{18}$O$^+$, H$^{13}$CN, and HNC are well correlated and all of them decrease with molecular hydrogen density, which seems to be one of the main factors determining their abundances in starless cores. This anti-correlation spans over around three orders of magnitude in density, from n(H$_2$)$\sim$ 10$^{3}$ cm$^{-3}$ to n(H$_2$)$\sim$ 10$^6$  cm$^{-3}$. 

\item Strong linear correlations are also found between abundances of the S-bearing species CS, SO, $^{34}$SO, H$_2$S, and OCS, which is very likely caused by the decrease of sulfur in gas phase. Under the reasonable assumption that the S/H in gas phase, (S/H)$_{gas}$, is correlated with the abundances of these species, we obtain that (S/H)$_{gas}$ $\propto$ n$^{-0.6}$ in the n(H$_2$)$\sim$ 10$^{3}-$10$^6$ cm$^{-3}$ density range.

\item The observed anti-correlation of the studied molecular abundances with density is always stronger in the case of the Taurus cloud. 
Perseus and Orion show weaker anti-correlations with H$_2$ density with smaller correlation parameters. However, we do recover strong anti-correlations with density for bins of 8$-$20 mag in Perseus and $>$20 mag in Orion. This shows that the dark region is located at a different visual extinction depending on the star formation activity.
 
\end{itemize}

In addition to the statistical analysis, we computed mean values of molecular column densities and abundances in Taurus, Perseus, and Orion in order to compare with extragalactic studies. The C$^{18}$O abundance is quite uniform in the three clouds, suggesting that its column density is a good gas mass tracer. However, the abundance of most species decreases, following the sequence Taurus, Perseus, and Orion, as a consequence of density distributions in these clouds.  

The GEMS project provides an unprecedented database with which to investigate the gas chemistry in molecular clouds and starless cores. This database is also useful to compare with molecular observations in external galaxies. A complete analysis of the chemical processes involved in the detected relations requires the comparison of observational data with predictions of chemical models, which will be carried out in a forthcoming paper. 

\begin{acknowledgements}
We thank the Spanish the Spanish Ministerio de Ciencia e Innovaci\'on for funding support through AYA2016-75066-C2-1/2-P and PID2019-106235GB-I00. 
SG-B acknowledges support through  grants PGC2018-094671-B-I00 (MICIU/AEI/FEDER,UE) and PID2019-106027GA-C44. 
JRG acknowledges support through  grants AYA2017-85111-P and PID2019-106110GB-I00.
I.J.-S. has received partial support from the Spanish FEDER (project number ESP2017-86582-C4-1-R) and the State Research Agency (AEI; project number PID2019-105552RB-C41). 
SPTM acknowledges to the European Union's Horizon 2020 research and innovation program for funding support given under grant agreement No~639459 (PROMISE).
We thank the anonymous referee for valuable comments that improved the manuscript.

\end{acknowledgements}

\bibliography{gems}

\end{document}